\documentclass[nohyper, letterpaper,11pt,notoc]{JHEP3}
\pdfoutput=1
\usepackage{amsfonts,amsmath,amssymb}
\usepackage{bm,bbm,url}
\usepackage{graphicx,epsfig}
\usepackage{xspace}

\newcommand{\gsim}{\mathrel{\lower0.8ex\vbox{\lineskip=0.15ex\baselineskip=0ex
                   \hbox{$>$}\hbox{$\sim$}}}}
\newcommand{\lsim}{\mathrel{\lower0.8ex\vbox{\lineskip=0.15ex\baselineskip=0ex
                   \hbox{$<$}\hbox{$\sim$}}}}

\newcommand{\sla}[1]{{\raise.15ex\hbox{$/$}\kern-.57em #1}}
\newcommand{\Sla}[1]{\kern0.12em{\raise.15ex\hbox{$/$}\kern-.74em #1}}
\newcommand{\del}{\partial}
\newcommand{\tr}{{\rm Tr}}
\newcommand{\wbar}[1]{\overline{#1}}
\newcommand{\wtild}[1]{\widetilde{#1}}
\newcommand{\dubdel}{\!\!\stackrel{\leftrightarrow}{\raise.02ex\hbox{$\del$}}}
\newcommand{\dubD}{\!\!\stackrel{\leftrightarrow}{\raise.02ex\hbox{$D$}}}

\newcommand{\beq}{\begin{eqnarray}}
\newcommand{\eeq}{\end{eqnarray}}
\newcommand{\nn}{\nonumber}

\newcommand{\varep}{\varepsilon}

\newcommand{\col}{{\tilde{\rho}}}
\newcommand{\ourpi}{{\tilde{\pi}}}

\newcommand{\Sherpa}{S\protect\scalebox{0.8}{HERPA}\xspace}
\newcommand{\Comix}{C\protect\scalebox{0.8}{OMIX}\xspace}
\newcommand{\Amegic}{A\protect\scalebox{0.8}{MEGIC++}\xspace}
\newcommand{\Pythia}{P\protect\scalebox{0.8}{YTHIA}\xspace}
\newcommand{\Alpgen}{A\protect\scalebox{0.8}{LPGEN}\xspace}
\newcommand{\PGS}{PGS\xspace}

\preprint{Edinburgh 2008/38}

\title{Searching for Multijet Resonances at the LHC}

\author{Can Kilic$^{(a)}$, Steffen Schumann$^{(b)}$ and Minho Son$^{(a)}$
\\ \it{(a) Department of Physics and Astronomy, Johns Hopkins University,\\3400 N. Charles St., Baltimore, MD 21218, USA}
\\ \it{(b) SUPA, School of Physics and Astronomy, The University of Edinburgh,\\JCMB Mayfield Road, Edinburgh, EH9 3JZ, UK\\
      (New address: Institut f{\"u}r Theoretische Physik, Universit{\"a}t Heidelberg,\\Philosophenweg 16, D-69120, Heidelberg, Germany)}}

\abstract{Recently it was shown that there is a class of models in which colored vector and scalar resonances can be copiously produced at the Tevatron
with decays to multijet final states, consistent with all experimental constraints and having strong discovery potential. We investigate the collider
phenomenology of TeV scale colored resonances at the LHC and demonstrate a strong discovery potential for the scalars with early data as well as the
vectors with additional statistics. We argue that the signal can be self-calibrating and using this fact we propose a search strategy which we show to be
robust to systematic errors typically expected from Monte Carlo background estimates. We model the resonances with a phenomenological Lagrangian that
describes them as bound states of colored vectorlike fermions due to new confining gauge interactions. However, the phenomenological Lagrangian treatment
is quite general and can represent other scenarios of microscopic physics as well.}

\keywords{Jets, Beyond Standard Model, Hadronic Colliders, Phenomenological Models}

\begin{document}
\section{Introduction}
\label{sec:intro}

Hadron colliders provide the highest center of mass energy in particle physics and are our portal to discovering heavy colored particles. We are currently
at the dawn of the LHC era where we hope to discover not only the Higgs boson but other states beyond the Standard Model (SM) as well. While the
``hierarchy problem'' has been a useful guide in approaching this critical time, one should be open to the possibility that the mechanism responsible for
the hierarchy may be beyond the reach of the LHC, and the data may present us with surprises that may or may not be related to this mechanism. From a more
general viewpoint, it is important to look for other principles to direct us towards the discovery of new physics in order to make sure that we can take
full advantage of the discovery potential of the LHC.

A general strategy in approaching the unknown is simply to utilize the main strength of one's tools, and in the case of hadron colliders this means the
production of colored states with large cross sections. With this in mind, it is surprising that there have been no dedicated searches for new physics in
the multijet channel in the past. This is even true for models of new physics that do have multijet final states, but since the overwhelming majority of
such models are aimed at solving the hierarchy problem they almost always contain additional final states that can be used to reduce background, such as
leptons or the presence of missing transverse energy. The obvious difficulty for a search in the multijet channel is the intimidating magnitude of the QCD
background, however it has recently been shown in ref.~\cite{Kilic:2008pm} that there is a broad class of models beyond the SM with minimal ingredients,
where the new physics would manifest itself only in the multijet channel but nevertheless the signal can easily compete with the background in terms of
cross section. These models only require the presence of new colored fermions (which already appear in almost all proposed extensions of the SM and are
essential for their discovery at hadron colliders) that are also charged under an additional gauge group which confines at energies moderately large
compared to the QCD scale. The bound states of the new gauge group, dubbed ``hypercolor'', can then be produced copiously at hadron colliders, their
existence is consistent with all experimental bounds even if they have masses as light as a few hundred GeV, and most importantly they can be discovered
with a dedicated resonance search in multijets. It was shown in ref.~\cite{Kilic:2008pm} that the Tevatron has a strong discovery potential for such
models at sub-TeV scales, and the aim of this paper is to provide a complimentary outlook for an LHC search in the TeV mass range.

When colored new fermions get bound into mesons of hypercolor, the state that corresponds to the rho meson of QCD is a color octet vector particle called
the coloron.\footnote{We borrow here the terminology of refs.~\cite{coloron-1} and \cite{coloron-2}.} The coloron can be resonantly produced at hadron
colliders as it can mix with the SM gluon. This process can microscopically be viewed as the pair creation of the new colored fermions which get bound by
the strong hypercolor interactions. In ref.~\cite{Kilic:2008pm} it was shown that if hypercolor is taken to be SU(3) with a set of fermions that transform
as fundamentals of SM color as well as hypercolor, then this model becomes a clone of QCD in terms of its ingredients, and one can use QCD as an analog
computer to sidestep the computational limitations of strong dynamics and quantitatively predict the spectrum of the hypercolor mesons. In fact, for this
benchmark model as well as for many other choices of the hypercolor gauge group and the representations of the matter fields, the coloron decays to a pair
of color octet pseudoscalar mesons of hypercolor, called hyperpions, which in turn decay to a pair of gluons, leading to events with a four jet resonance.
These states also appear in non-minimal Technicolor theories, see for example refs.~\cite{Farhi:1979zx,Eichten:1984eu,Lane:2002sm}.

In this paper we will begin by retracing the analysis of ref.~\cite{Kilic:2008pm} for the LHC, where the pair production of hyperpions through their
minimal QCD gauge coupling becomes more significant than their resonant production through the coloron, due to the dominance of gluon PDF's at the LHC
over quark-antiquark PDF's. While the strong evidence for the secondary resonances is unaffected, this can weaken the evidence for the presence of the
primary resonance, the coloron. Therefore we will extend the analysis of collider phenomenology for such models by including the process of coloron pair
production, which leads to eight jet final states. Even though current Monte Carlo based methods can be unreliable in estimating detailed kinematic
features of high multiplicity multijet backgrounds, we will argue that signal, where the primary production process has a two body final state, has a
large enough cross section to overcome background involving the QCD production of hard and well separated multi-parton final states, even without taking
advantage of the detailed kinematic features. In addition, we will demonstrate that signal events can be reconstructed to reveal kinematic features which
have very low probability to be present in background events, thus leading to a further very significant enhancement of signal over background and making
the coloron discoverable. For the sake of completeness we will report the results of applying our reconstruction procedure on a background sample that is
plausibly realistic, thereby providing a check on our conclusions.

The clear invariant mass peak in hyperpion pair production and the large signal cross section for coloron pair production also allows for a simple search
strategy where the signal can be self calibrating. We will propose a guideline for such a search where the coloron mass scale can manifest itself as an
anomaly in the $h_{T}$ distribution after requiring eight hard jets. We will argue that this procedure can work even allowing for relatively large
systematic errors on the background. This measurement can be used to optimize the cuts for the subsequent part of the analysis for coloron mass
reconstruction.

The paper is organized as follows: In section \ref{sec:model} we will outline the general scenarios of new physics we are interested in that lead to
multijet resonances at the LHC, and specify a phenomenological Lagrangian that describes the production and decay of the hypermesons. The parameters of
this phenomenological Lagrangian are fundamentally determined by the choice of the hypercolor interactions and the transformation properties of the new
matter fields under SM color and hypercolor. We will then concentrate on a representative choice for these parameters and use this benchmark model to show
in section \ref{sec:search} that the LHC has strong discovery potential for the scenarios at hand, with a relatively small amount of data for a wide range
of mass scales. We will demonstrate that combined evidence from four jet and eight jet final states can conclusively point to the new physics scenarios
described above. Finally, we will outline a search strategy where signal can be self calibrating such that each stage of the analysis determines the
optimized cuts to be used at the next stage. We will present our conclusions in section \ref{sec:conclusions}.

\section{A Quantitative Model}
\label{sec:model}

The scenarios we wish to discuss require a minimal set of ingredients in addition to the SM, the Lagrangian can be written down as
\beq
  {\cal L}
  = {\cal L}_{\rm SM} + \bar{\psi} (i \Sla{D} - m) \psi
   -\frac{1}{4} H_{\mu \nu} H^{\mu\nu}  \>.
\label{eq:UV-Lag} \eeq
where the $\psi$ are fermions charged under $SU(3)_{c}$ as well as the hypercolor gauge group (represented by $H_{\mu \nu}$). Despite the simplicity of
this setup, the resulting phenomenology is rich, and was discussed in detail in ref.~\cite{Kilic:2008pm}. While the choices for the hypercolor group and
the group representations that the $\psi$ transform under are manifold, certain qualitative features of the spectrum below $\Lambda_{HC}$, the hypercolor
confinement scale are common. For instance, the existence of an $SU(3)_c$ adjoint vector $\col_{\mu}$ (the coloron) and the parametrically lighter
$SU(3)_c$ adjoint scalar $\ourpi$ (hyperpion) are robust (we use tildes throughout the paper to distinguish bound states of hypercolor from those of QCD).
Even though we can say little else in quantitative detail about many of the possible choices, specifying the $\psi$ to be $SU(3)_c$ triplet fermions,
these states are the only ones below $\Lambda_{HC}$ relevant for collider phenomenology. Below, we write down a phenomenological Lagrangian that captures
the interactions of these states as well as their couplings to the SM, expanding upon the one presented in ref.~\cite{Kilic:2008pm} by adding two terms
relevant for the LHC phenomenology that were unimportant for the Tevatron. \footnote{We assume here that the coloron and gluon kinetic terms have already
been diagonalized, for further details see ref.~\cite{Kilic:2008pm}.}:
\beq
  {\cal L}_{\rm eff}^{\rm HC}
  &=& -\frac{1}{4} G^{a}_{\mu\nu} G^{a\mu\nu}+\wbar{q} i\gamma^{\mu}\left(\partial_{\mu}+ig_{3}\left(G_{\mu}+\varep\col_{\mu}\right)\right)q\nn\\
  && -\frac{1}{4}\left(D_{\mu}\col_{\nu}-D_{\nu}\col_{\mu}\right)^{a}\left(D^{\mu}\col^{\nu}-D^{\nu}\col^{\mu}\right)^{a}+\frac{m_{\col}^{2}}{2}\col^{a}_{\mu}\col^{a\mu}\nn\\
  && +i\chi g_{3}\tr\left(G_{\mu\nu}\left[\col^{\mu},\col^{\nu}\right]\right)+\xi\frac{2i\alpha_{s}\sqrt{N_{HC}}}{m^{2}_{\col}}\ \tr\left(\left(D^{\mu}\col_{\nu}-D_{\nu}\col^{\mu}\right)\left[G^{\ \nu}_{\sigma},G^{\ \sigma}_{\mu}\right]\right)\label{eq:eff-Lag}\\
  && +\frac{1}{2} (D_{\mu}\ourpi)^{a}(D^{\mu}\ourpi)^{a}-\frac{m_{\ourpi}^{2}}{2}\ourpi^{a}\ourpi^{a}-g_{\col\ourpi\ourpi}f^{abc}\col^{a}_{\mu}\ourpi^{b}\del^{\mu}\ourpi^{c}-\frac{3g_{3}^{2}}{16\pi^{2}f_{\ourpi}}\tr\bigl[\ourpi G_{\mu\nu}\wtild{G}^{\mu\nu}\bigr]\>\nn\>.
\eeq

The first line contains the SM kinetic terms as well as the coupling of the coloron to the quarks, parameterized by the effective mixing parameter
$\varep$. The second line contains the quadratic coloron terms, where $D_{\mu}$ stands for the $SU(3)_c$ covariant derivative acting on an adjoint. Note
that both the coloron and hyperpions have minimal gauge couplings to gluons due to their color adjoint nature, this allows them to be pair produced from a
$g$-$g$ initial state, which we will refer to as ``pair production through QCD''. The terms on the third line were not included in
ref.~\cite{Kilic:2008pm} as they were not relevant for the Tevatron phenomenology, they are important for the LHC however and we include them here (with
the $SU(3)$ generators normalized as $\rm{Tr}\left(T^{a}T^{b}\right)=\frac{1}{2}\delta^{ab}$). These terms represent strong interaction matrix elements of
the underlying theory which cannot be extracted from the QCD analogy, we do however write them down making their natural sizes manifest in the spirit of
naive dimensional analysis (NDA) \cite{Manohar:1983md,Georgi:1985kw} with free coefficients of order one. The first term of the third line is
renormalizable and is gauge invariant by itself, with a coefficient $\chi$, we will say more about this shortly. The second term of the third line is
non-renormalizable, however it is the leading term that leads to resonant coloron production from two incoming gluons, and since the $q$-$\bar{q}$ PDF's
are not as dominant at the LHC as at the Tevatron, we choose to keep this term. We estimate the size of its coefficient using NDA up to an ${\cal{O}}(1)$
number, which we denote by $\xi$. The last line includes the kinetic terms for the hyperpion, the effective vertex that leads to the dominant coloron
decay mode into a pair of hyperpions parameterized by $g_{\col\ourpi\ourpi}$ as well as the axial anomaly term that leads to the hyperpion decay into two
gluons.

At this point the effective parameters appearing in eq.~(\ref{eq:eff-Lag}) are undetermined. Although some of them are in principle calculable on the
lattice, only little is known for generic choices of the hypercolor group or the hypercolor representation of the $\psi$.\footnote{For a first attempt to
simulate $SU(N)$ gauge theories for arbitrary fermion representations and generic numbers of colors $N$, and in particular $SU(2)$ with adjoint fermions,
see ref.~\cite{DelDebbio:2008zf} and references therein.} There is a special case however, where almost all parameters can in fact be determined
quantitatively. This special choice corresponds to the $SU(3)$ hypercolor gauge group, with the $\psi$ transforming in the fundamental representation.
Then the zeroth order hypercolor dynamics below $\Lambda_{HC}$ (treating QCD as a perturbation on hypercolor) are identical to the zeroth order QCD
dynamics below $\Lambda_{QCD}$ (treating QED and quark masses as a perturbation on QCD). In the remainder of the paper we will refer to this choice as the
benchmark model. Ref.~\cite{Kilic:2008pm} extracted the effective parameters of the phenomenological Lagrangian in eq.~(\ref{eq:eff-Lag}) in the benchmark
model by scaling up the QCD spectrum and the effective couplings of the QCD chiral Lagrangian to obtain
\beq
\varep&\simeq&0.2\,,\\
g_{\col\ourpi\ourpi}&\simeq&6\,,\\
\frac{m_{\ourpi}}{m_{\col}}&\simeq&0.3\,,\label{eq:mrho_BM}\\
\frac{f_{\ourpi}}{\Lambda_{HC}}&\simeq&\frac{f_{\pi}}{\Lambda_{QCD}}\,. \eeq

While the benchmark model may appear to be a very specific choice for the hypercolor sector, its collider phenomenology is actually representative of a
wide range of choices. In fact, the main effect of the choice of the hypercolor group and the representation to which the $\psi$ belong (encoded in the
effective parameters in eq.~(\ref{eq:eff-Lag})) can be represented by a small number of independent quantities, namely the masses of the coloron and the
hyperpions, the production cross section of the coloron and its branching fraction into dijets vs. four jets (via hyperpions). It is therefore useful to
describe the collider phenomenology in terms of these quantities and we will now elaborate on how the phenomenology is affected as they are varied away
from the benchmark model. For values of $\frac{\varep}{g_{\col\ourpi\ourpi}}$ or $\frac{m_{\ourpi}}{m_{\col}}$ that are too large, the hyperpion decay
mode may be suppressed or kinematically inaccessible and dijets will be the dominant decay mode, on the other hand if the value of
$\frac{m_{\ourpi}}{m_{\col}}$ is very small, then the hyperpions from the coloron decay will be very boosted and this can lead to the two jets from each
hyperpion to be reconstructed as a single jet, thereby also turning the phenomenology effectively into the case where dijets are the dominant decay mode.
It was shown in ref.~\cite{Kilic:2008pm} that the main constraint on the models at hand comes from resonance searches in the dijet channel, and colorons
which decay dominantly into dijets are excluded at sub-TeV scales \cite{Abe:1997hm,Abazov:2003tj,Hatakeyama:2008tz}.\footnote{A range of other potential
constraints on this type of new physics were discussed in ref.~\cite{Kilic:2008pm} and shown to be irrelevant even at sub-TeV scales. We will not repeat
this analysis here as we will be interested in TeV-scale masses where the constraints are even weaker.} On the other hand, as long as one stays away from
these extreme regions the model is consistent with all present constraints and the collider phenomenology is virtually unchanged. Furthermore, above
$m_{\col}\sim1~\rm{TeV}$ even a coloron decaying predominantly into dijets is not excluded by Tevatron dijet resonance searches, however we will not
discuss this possibility in detail since there are already dedicated searches for dijet resonances proposed for the LHC
\cite{Cardaci:2008vg,Bhatti:2008hz}. We will use the parameters of the benchmark model to generate Monte Carlo signal events in section \ref{sec:search}
simply because this is a representative choice, however our analysis strategy will be blind, and free of any assumptions regarding the values of the
effective parameters.

As mentioned, the value of the parameter $\chi$ cannot be obtained from the QCD analogy (this term vanishes for QCD when $G_{\mu\nu}$ is replaced by
$F_{\mu\nu}^{em}$), however NDA predicts its natural size to be $\mathcal{O}(1)$. In fact, we can cross check this by considering similar couplings that
appear in other models of new physics containing a massive spin-1 color octet state, such as a two-site model where the $SU(3)_{L}\times SU(3)_{R}$ is
broken by the vev of a scalar bifundamental and SM color is the unbroken $SU(3)$ (The heavy vector adjoint in such models is known as an axigluon
\cite{Frampton:1987dn}) or the case of five dimensional QCD where the extra dimension has been compactified on a circle. In both cases the value of $\chi$
is fixed at $\chi=1$. In the case of 5-D QCD which is perturbative in the UV, this can be traced back to 5-D Lorentz invariance, and in fact $\chi=1$ is
the unique choice for which the scattering amplitude for $G^{(0)}G^{(0)}\rightarrow G^{(1)}G^{(1)}$ remains unitary at high energies. As we will see in
section \ref{sec:search}, the main effect of $\chi$ will be on the coloron pair production cross section, and we will adopt the choice $\chi=1$ for
generating signal events.

The parameter $\xi$ is another undetermined number of order one, and while we will demonstrate
how the cross section for hyperpion pair production depends on $\xi$, we will stick to the choice
of $\xi=0$ in generating signal events, as this is the most conservative choice (non-zero values
of $\xi$ enhance the cross section). In terms of coloron decay branching fractions, a nonzero
value of $\xi$ gives rise to a two-gluon decay mode, but the branching fraction is quite small,
about $7.5\times10^{-3}$ for $\xi=1$ so the collider phenomenology is unaffected unless $\xi$
is much larger than its natural size.

\section{Search Channels and Strategies}
\label{sec:search}

\FIGURE[!t]{\includegraphics[width=5.0in]{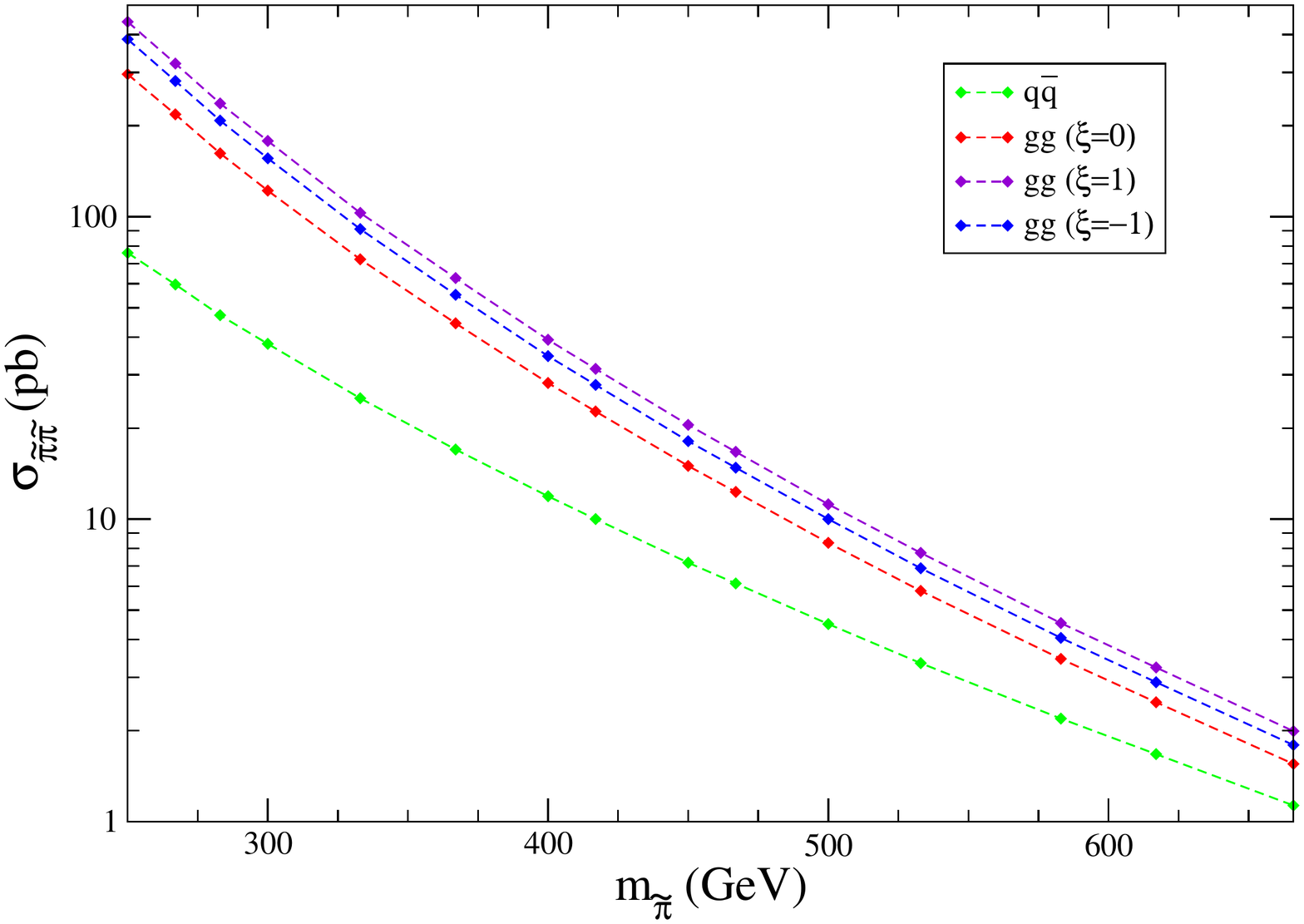} \caption{Hyperpion pair production cross section at the LHC, assuming the coloron mass is
given through eq.~(\ref{eq:mrho_BM}). The green curve represents the resonant coloron production cross section which subsequently decays to a pair of
hyperpions, while the red curve represents the pair production cross section of hyperpions from two initial gluons. The blue and violet curves demonstrate
the enhancement of the gluon initial state cross section as $\xi$ is allowed to be as large as its natural size.} \label{fig:xsecpipi}}
The most obvious place to search for the new states is the resonant production of the coloron. This channel has strong discovery potential at the Tevatron
for masses accessible there \cite{Kilic:2008pm}, however there are qualitative differences for the LHC. The most straightforward way to resonantly produce
a coloron is through a $q$-$\bar{q}$ initial state, however the PDF's at the LHC suppress the resonant cross section in comparison to the Tevatron. While
there is also the potential of resonant coloron production from a two gluon initial state, the cross section for hyperpion pair production is actually
dominated by the minimal QCD coupling of hyperpions to gluons. This is displayed in figure \ref{fig:xsecpipi}, where the green curve represents the
hyperpion pair production cross section coming from the coloron resonance, while the red curve represents the hyperpion pair production cross section from
initial state gluons, with $\xi=0$. The blue and violet curves demonstrate the effect of turning on a nonzero value for $\xi$, which is to increase the
cross section. The most important consequence is that even if we can isolate signal events by the fact that they contain two resonances with equal masses,
in general we will not be able to reconstruct a primary resonance. Thus at the LHC the coloron is more difficult to discover compared to the hyperpions.
After describing in section \ref{subsec:simulation} the tools used in our analysis, we will demonstrate in section \ref{subsec:4j} the discovery potential
for hyperpions in the four jet channel, then in section \ref{subsec:8j} we will augment our analysis by a search for pair produced colorons in an eight
jet channel. We will demonstrate the range of masses where the LHC has a strong discovery potential by working with two cases, $m_{\col}=750~\rm{GeV}$
(below which the coloron is discoverable at the Tevatron \cite{Kilic:2008pm}) and $m_{\col}=1.5~\rm{TeV}$ (above which in the benchmark model the coloron
pair production cross section becomes too small, even though there is still significant room for the discovery of the hyperpions).

Monte Carlo based methods for background estimates are expected to suffer from systematic errors at high parton multiplicity, and we will take care in
making statements about the eight-jet channel conservatively, using only cross section information for background (with expected systematic errors of
order one) using the most sophisticated Monte Carlo tools available for high multiplicity. We will argue that the large signal cross section will
allow us to draw conclusions that should remain valid even in the presence of sizeable corrections to our background estimates. We will demonstrate in appendix \ref{app:8j-bg} that our conclusions are indeed valid when applied to a plausibly realistic sample of background events, generated
with our limitations of available computing power.

Section \ref{subsec:generic search} will be devoted to showing how this type of signal can be self calibrating and proposing a search strategy where one
can extract the optimized cuts for the eight jet analysis using the mass peaks in the four jet analysis and the data at high $h_{T}$. This procedure is
summarized visually in figure \ref{fig:flowchart}, while we will argue the success of the four and eight jet analyses are robust, the extraction of the
$h_{T}$ peak may break down for systematic errors of $\mathcal{O}(10)$ on the magnitude of our background estimates. The second step however should be
regarded as optional and should not impact the discoverability of the hyperpions or the coloron.

\FIGURE[!t]{\includegraphics[width=4.0in]{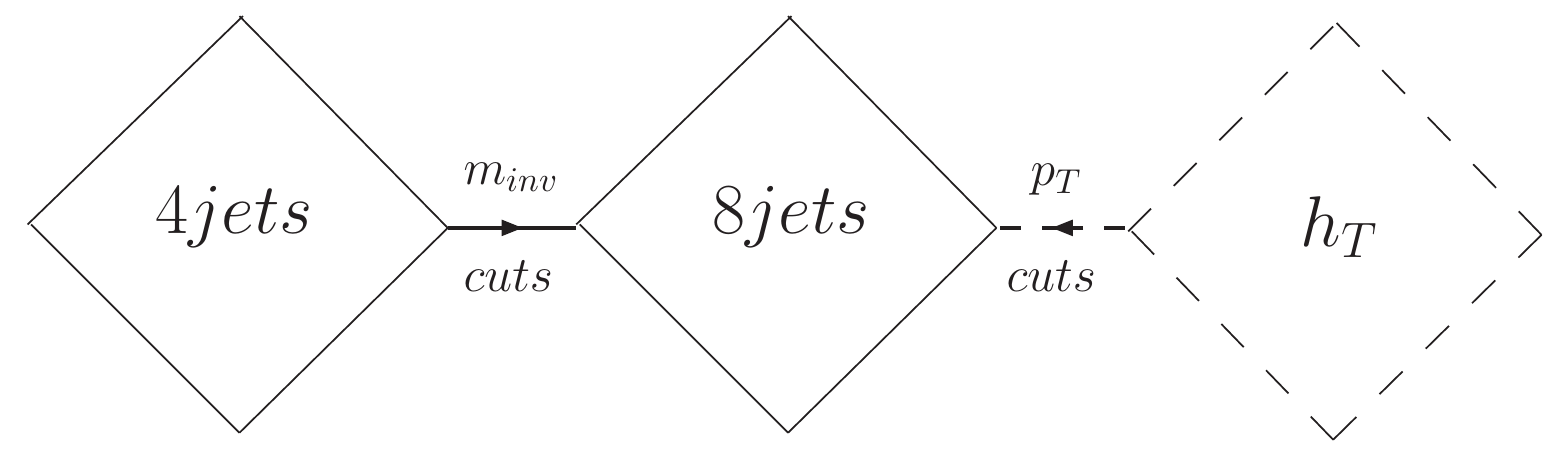} \caption{A visual representation of the search strategy we propose. While the extraction
of the $h_{T}$ peak may become unrealistic if the systematic errors of our background estimates are $\mathcal{O}(10)$, the success of the rest of the
analysis should be robust.}\label{fig:flowchart}}

\subsection{Simulation of Signals and Backgrounds}
\label{subsec:simulation}

Before discussing in detail the individual search channels and establishing the methods to extract signal from background we want to describe in detail
the tools we used to simulate the signal processes and QCD multijet production.

\subsubsection{Signal simulation}
\label{subsubsec:signal_simulation} For the calculation of signal cross sections and the generation of signal events at parton level we have incorporated
the phenomenological Lagrangian presented in eq.~(\ref{eq:eff-Lag}) in the \Sherpa Monte Carlo program \cite{Gleisberg:2003xi} by implementing all
interaction vertices of the model into \Sherpa's matrix element generator \Amegic \cite{Krauss:2001iv}, thereby allowing us to generate arbitrary
tree-level processes within this model. We limit the use of full matrix-element calculations to the point of production of the hyperpions thereby
capturing non-resonant QCD contributions as well as finite-width effects for the intermediate heavy states. The latter is of particular importance when
studying coloron production as for our benchmark model the vector states are rather broad. In fact the ratio of $\Gamma_{\col}/m_{\col}\approx 0.2$
prohibits the use of a simple narrow-width-approximation (NWA) for the coloron decays. Furthermore, using exact matrix element calculations,
spin-correlations between the coloron decay products are properly taken into account. For the decays of the hyperpions themselves we use a NWA
($\Gamma_{\ourpi}/m_{\ourpi}\approx 2\times10^{-4}$) and distribute the decay gluons isotropically in phase space consistent with the decay of a scalar
intermediate state. The parton-level computations are done using the CTEQ6L parton-distribution functions \cite{Pumplin:2002vw}. The factorization and
renormalization scales entering are set equal to the mass of the pair-produced states, namely $\mu_F=\mu_R=m^2_{\tilde \pi}$ for hyperpion pair production
and $\mu_F=\mu_R=m^2_{\tilde \rho}$ for the coloron-pair channel. To account for additional jet radiation and intra-jet evolution the parton-level events
are passed to \Pythia (version 6.4) \cite{Sjostrand:2003wg} for parton showering and subsequent hadronization. Hadron-level events are processed through
\PGS (version 4.0) \cite{pgs} to include effects of finite detector resolution as well as jet reconstruction. We use a cone jet algorithm with $R=0.5$
throughout our analysis.

\subsubsection{Background simulation}
\label{subsubsec:background_simulation}

Simulating the Standard Model backgrounds for hyperpion and coloron pair-production implies generating inclusive four- and eight jet production,
respectively. For the signal processes we expect rather hard jets from the decays of the heavy resonances, therefore we will always use cuts to demand
high $p_{T}$ jets. Realistically estimating the rates for producing several well-separated and hard jets requires the use of a complete matrix-element
calculation rather than the parton-shower approach. Lacking full one-loop QCD results we have to rely on tree-level calculations of the corresponding
amplitudes. Intrinsically these computations suffer from significant scale uncertainties that affect both total cross sections and kinematical
distributions. However, in our four jet analysis we will demonstrate very clearly that the hyperpion can be reconstructed such that the resonance is
unmistakable even allowing for order one errors on the Monte Carlo background estimates. Coloron pair production, which is a $2\rightarrow 2$ process, has
a large enough cross section to overcome the $2\rightarrow 8$ background even without taking advantage of detailed kinematic features of the signal. We will further show
that signal events can be reconstructed to pass very nontrivial kinematic cuts which further reduces background by a very significant amount.

To compute the cross section of the four- and eight jet channels the new matrix-element generator \Comix \cite{Gleisberg:2008fv}, interfaced with the
\Sherpa event-generation framework, is used. \Comix constructs the QCD amplitudes using color-dressed Berends--Giele recursion \cite{Duhr:2006iq} rather
than traditional Feynman diagrams. To integrate the multi-parton phase space \Comix builds suitable phase-space mappings that are managed using an
adaptive multi-channel method \cite{Gleisberg:2008fv}. The combination of these techniques makes \Comix very efficient and fast and therefore well suited
for our purposes. For all parton-level background calculations we use a dynamical factorization and renormalization scale, namely the average
transverse-momentum squared of all final-state partons. In Tab.~\ref{tab:4jet_PLxs} we present four jet cross sections for two jet-$p_{T}$ thresholds,
namely $p_{T,j}>100$ GeV and $p_{T,j}>200$ GeV. Furthermore, all partons are required to satisfy
\begin{equation}\label{eq:4jet_PLcuts}
\Delta R_{jj}>0.5\quad {\rm and}\quad |\eta_j|<2.0\,.
\end{equation}
\TABLE[!t]{
\begin{tabular}{|c||c|c|}\hline
$\mu_F= \mu_R$ & $\sigma_{4j}(p_{T,j}>100\,{\rm GeV})$ [pb] &
$\sigma_{4j}(p_{T,j}>200\,{\rm GeV})$ [pb] \\\hline
$\langle p^2_{T,j}\rangle$ & 1463(13) & 37.1(3) \\
$\frac14\langle p^2_{T,j}\rangle$ & 2404(24) & 58.5(6)\\
$4\langle p^2_{T,j}\rangle$ & 995(9) & 24.7(2)\\\hline
\end{tabular}
\caption{\label{tab:4jet_PLxs}LHC four jet production cross sections for jet-transverse-momentum thresholds of $100$ GeV and $200$ GeV. The central
factorization and renormalization scales are given by the averaged transverse-momentum squared of the final-state jets. The phase-space cuts given in
eq.~(\ref{eq:4jet_PLcuts}) have been imposed. The numbers in parentheses indicate the absolute statistical error on the last digit.}
}
To estimate the scale uncertainty of the cross sections we consider variations of the factorization and renormalization scales by common factors of $1/4$
and $4$, corresponding numbers are also given in Tab.~\ref{tab:4jet_PLxs}. The scale dependence of the cross sections can be estimated to about $\pm50\%$
and is dominated by the high power of the strong coupling entering the calculations. All results have been confirmed with \Alpgen
\cite{Mangano:2002ea}\footnote{However, in \Alpgen not all partonic subprocesses contributing to four jet production are available. Therefore a tuned
comparison has been performed and processes with more than two quarks in the final state were excluded in \Comix as well. The contribution of these
color-suppressed channels is less than $10\%$ of the total four jet cross sections.}.

When considering eight jet production there is an enormous number of partonic subprocesses that contribute, however, most of them contain multiple pairs
of quarks. Compared to the purely gluonic process these configurations are color-suppressed. To simplify our background calculations we only include
processes with at most two quarks in the initial and final state --- all the remaining partons being gluons. The complete list of channels we take into
account is
\begin{equation}\label{eq:8jet_procs}
gg \to 8g\,,\quad gg \to 6g2q\,,\quad gq \to 7g1q\,,\quad qq \to 8g\,,\quad qq \to 6g2q\,,
\end{equation}
where $q$ denotes all combinations of light-flavor quarks and anti-quarks allowed by QCD. As explicitly shown in ref.~\cite{Gleisberg:2008fv} this
captures the bulk of the total eight jet cross section and we estimate the contribution of the neglected channels to be less than $20\%$ of the total
cross section. In Tab.~\ref{tab:8jet_PLxs} we list corresponding results for eight jet production at the LHC both for our default scale choice and
variations of that by factors of $1/4$ and $4$. The phase-space cuts used for this channel are again given by eq.~(\ref{eq:4jet_PLcuts}), supplemented
with $p_{T}$ thresholds $p_{T,j}>100$ GeV and $p_{T,j}>200$ GeV. For these particular setups the scale variations change the results by approximately a
factor of two up and down. This can largely be ascribed to the renormalization scale dependence of the ${\cal{O}}(\alpha_S^8)$ process.
\TABLE[!t]{
\begin{tabular}{|c||c|c|}\hline
$\mu_F= \mu_R$ & $\sigma_{8j}(p_{T,j}>100\,{\rm GeV})$ [fb] & $\sigma_{8j}(p_{T,j}>200\,{\rm GeV})$ [fb] \\\hline
$\langle p^2_{T,j}\rangle$ & 112(1) & 0.369(5) \\
$\frac14\langle p^2_{T,j}\rangle$ & 267(3) & 0.848(9)\\
$4\langle p^2_{T,j}\rangle$ & 53.4(7) & 0.174(2)\\\hline
\end{tabular}
\caption{\label{tab:8jet_PLxs}LHC eight jet production cross sections for $p_{T,j}$ thresholds of $100$ GeV and $200$ GeV. Note that only the constrained
set of processes listed in eq.~(\ref{eq:8jet_procs}) is considered. The applied phase-space cuts are given in eq.~(\ref{eq:4jet_PLcuts}).}
}

For both signal and background, the parton-level events used in the four jet analysis have been interfaced to \Pythia and \PGS to account for parton
showering and hadronization, as well as detector effects. In the eight jet case we only use total integrated cross sections to quantify background (except
for the $h_{T}$ distributions presented in section \ref{subsec:generic search}) as we wish to avoid relying on detailed kinematic features of Monte Carlo
based eight jet background events. Furthermore, the unweighting efficiency for eight parton phase space is rather low which makes the generation of large
enough background event samples computationally very expensive. Once we demonstrate that the eight jet cross section for signal is larger than that of
background at parton level after using simple $p_{T}$ cuts, we will proceed with a signal only study for mass reconstruction (after passing signal events
through \Pythia and \PGS), which is justified since the small expected efficiency for background events to pass our reconstruction cuts should provide a
nearly pure signal sample. To corroborate our statements about the detectability of coloron pair production in the eight jet channel we
will confront the signal with a sample of showered and hadronized 6-parton background events. The corresponding analysis can be found
in appendix \ref{app:8j-bg}.

\subsection{Four Jet Analysis}
\label{subsec:4j}

In order to improve signal over background, we take advantage of the main kinematic difference between the two, namely that the jets in the signal events
come from the decay of two equal-mass resonances and are therefore likely to have comparable and large $p_{T}$'s while the jets in the background are
produced from QCD where the bulk of the background exhibits a strong $p_{T}$ hierarchy. Therefore for event selection, besides the generic cuts given in
eq.~(\ref{eq:4jet_PLcuts}), we impose a hard $p_{T}$ cut on the hardest four jets in each event.

We then pair the four hardest jets in all possible ways and discard the event if there is no pairing in which the invariant masses of the two pairs are
within $50~\rm{GeV}$ of each other. If there is more than one such pairing, we take the one that yields the closest masses for the pairs. We then take the
mean value of the pair masses as the candidate hyperpion mass, and we also plot the correlation between the pair mass and the total invariant mass of the
four jets. This separates signal from the bulk of the background because of the following reason: The background events that yield comparable pair masses
have a special kinematic configuration, namely the jets that make up the pair that does not contain the hardest jet are quite far apart in $\Delta R$ in
order to be able to match the invariant mass of the pair containing the hardest jet. This means that for a given value of the four jet invariant mass, the
background events typically have a larger value for the pair invariant mass compared to signal events. This can be seen clearly in figures
\ref{fig:pipi750} and \ref{fig:pipi1500}.

\subsubsection{Lighter mass case}

\FIGURE[!t]{\includegraphics[width=4.0in]{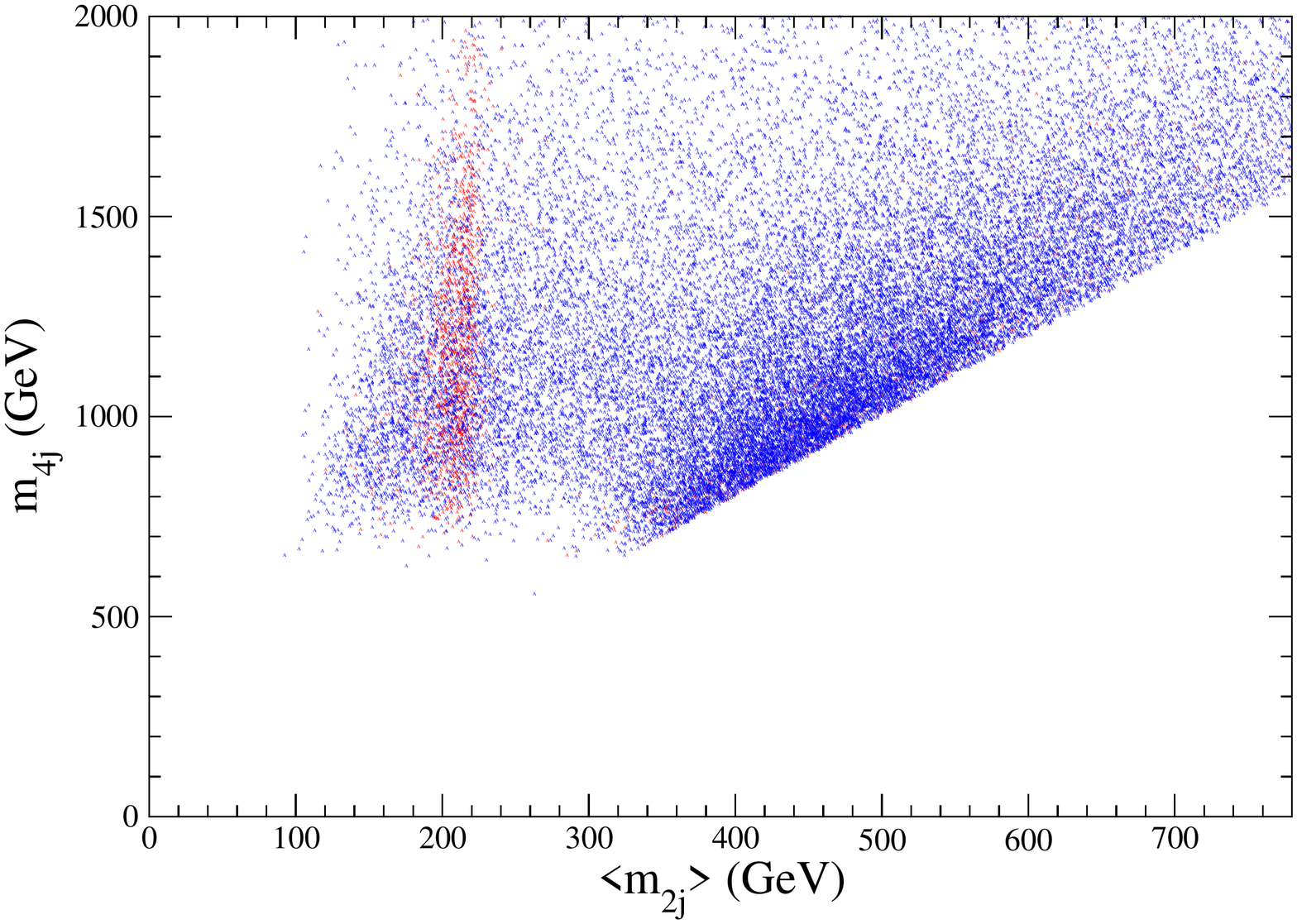} \includegraphics[width=4.0in]{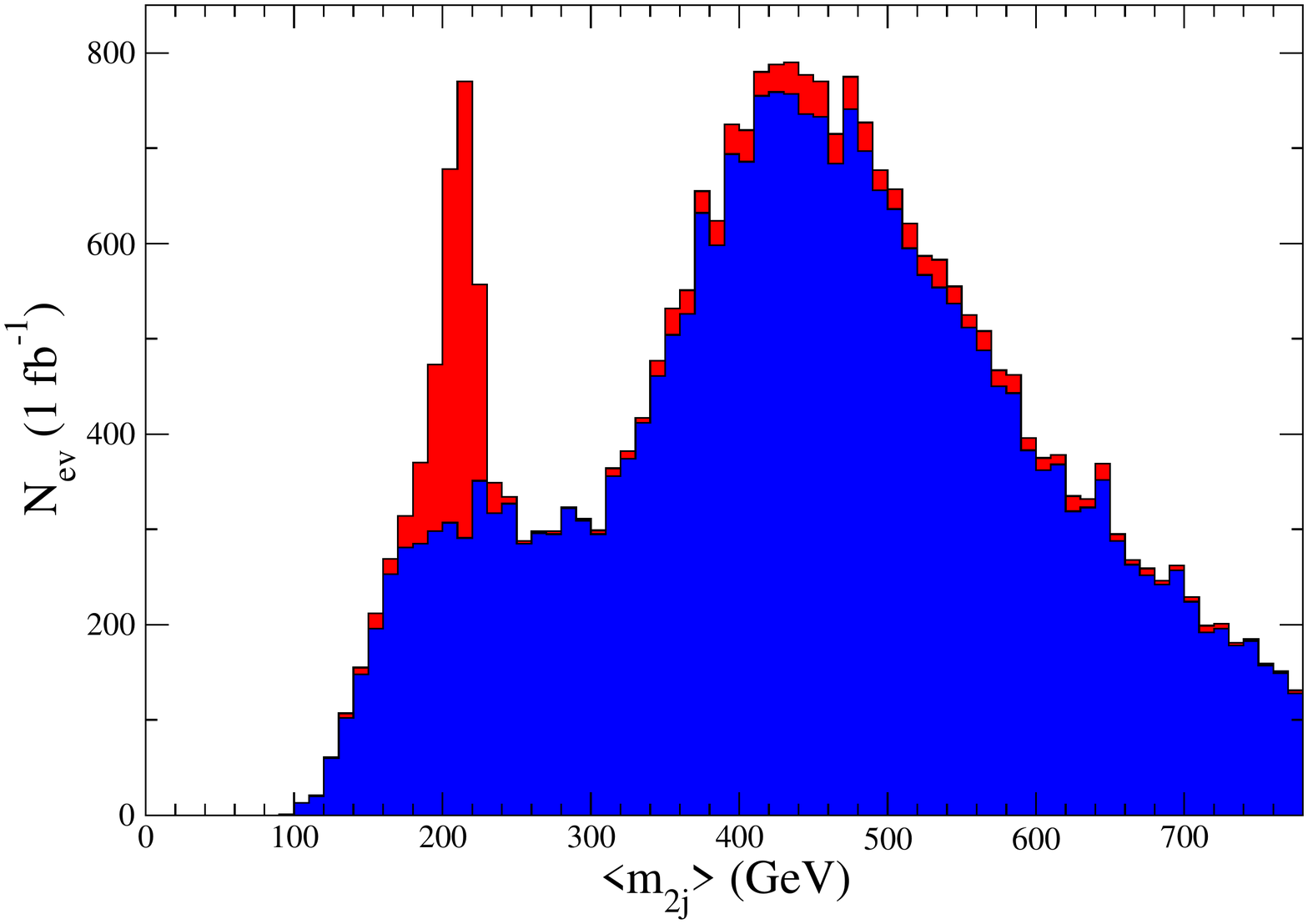} \caption{Upper plot: Pair invariant mass vs. four
jet invariant mass for $1~\rm{fb}^{-1}$ of data, red is signal while blue is background. The coloron mass is $750~\rm{GeV}$ and the hyperpion mass is
$225~\rm{GeV}$. Lower plot: The projection onto the pair invariant mass.} \label{fig:pipi750}}
For $m_{\col}=750~\rm{GeV}$ (and $m_{\ourpi}=225~\rm{GeV}$) we choose a $p_{T}$ cut of $150~\rm{GeV}$ on the four hardest jets. After the pairing and
invariant mass cuts, the signal cross section is $2.2~\rm{pb}$ and the background cross section is $24~\rm{pb}$. In figure \ref{fig:pipi750} one can see
that even though the pair mass has a peak at the hyperpion mass, the four jet mass values do not accumulate at the coloron mass, that is because in most
of the signal events hyperpion pairs are created through their gauge couplings to gluons.

We define the statistical significance in a histograms as
\begin{equation}
\chi^{2}_{\rm sig}=\sum_{\rm bins}\left(\frac{N_{S,{\rm bin}}}{\sqrt{N_{B,{\rm bin}}}}\right)^{2},
\end{equation}
\FIGURE[!t]{\includegraphics[width=4.0in]{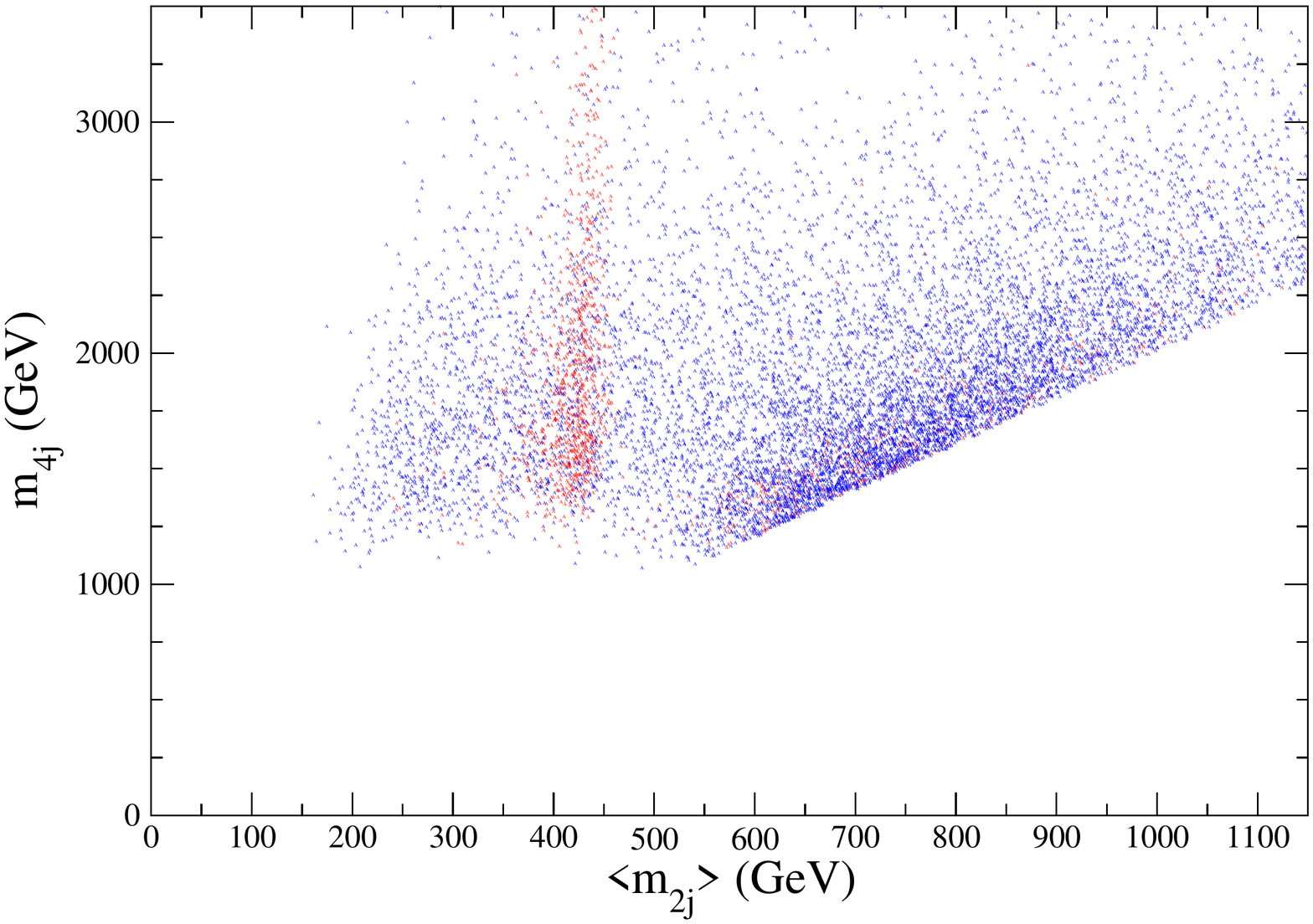} \includegraphics[width=4.0in]{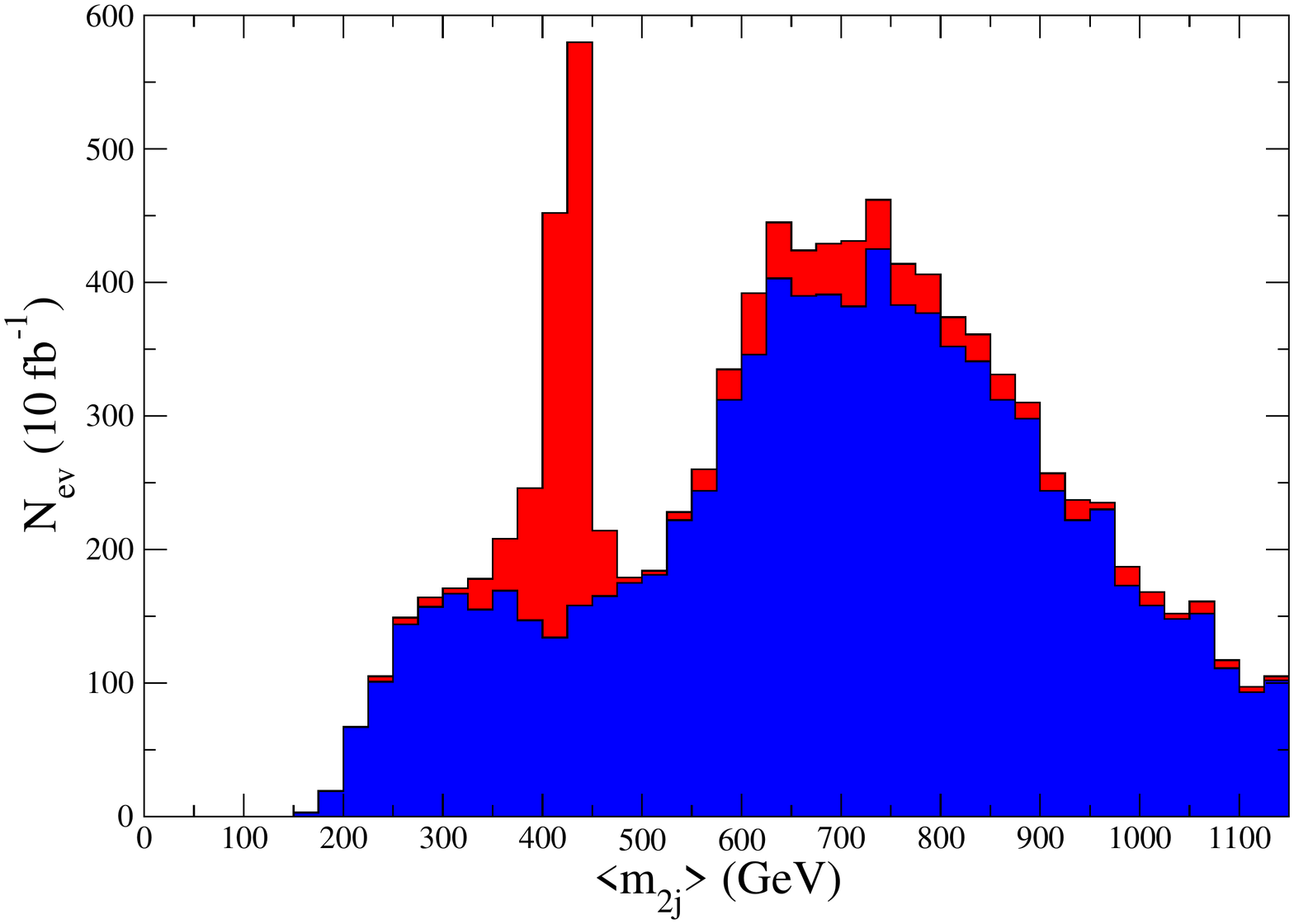} \caption{Upper plot: Pair invariant mass vs. four
jet invariant mass for $10~\rm{fb}^{-1}$ of data, red is signal while blue is background. The coloron mass is $1.5~\rm{TeV}$ and the hyperpion mass is
$450~\rm{GeV}$. Lower plot: The projection onto the pair invariant mass.} \label{fig:pipi1500}}
where the bin size is chosen such that each bin is reasonably well populated while any significant shapes in the distributions are retained. We check that
$\chi_{\rm sig}$ is invariant under $\mathcal{O}(1)$ changes in the bin size. For $1~\rm{fb}^{-1}$ of data the statistical significance of the histogram
in figure \ref{fig:pipi750} is $\chi_{sig}\sim 38$.

Figure \ref{fig:pipi750} should also alleviate most concerns about systematic errors in using Monte Carlo backgrounds. As the $p_{T}$ cut that goes into
this analysis is changed, the effect on the signal and background are quite different. The background distribution, in particular the position of the
artificial peak introduced by the $p_{T}$ cuts, shifts as the cut value is changed. The signal on the other hand is very clearly peaked at the true mass
of the hyperpion and only the height of the peak is affected as the $p_{T}$ cut is varied, since the number of events passing the cuts changes. Therefore
even allowing for errors of order one on the background, there will be a wide range for the value of the $p_{T}$ cut where the narrow resonant shape of
the signal will be clearly distinguishable from the background.

\subsubsection{Heavier mass case}
For $m_{\col}=1.5~\rm{TeV}$ (and $m_{\ourpi}=450~\rm{GeV}$) we impose a $p_{T}$ cut of $250~\rm{GeV}$ on the four hardest jets. After the $p_{T}$ cuts and
the pair invariant mass cut the cross section of signal is $0.15~\rm{pb}$ while the background cross section is $1.0~\rm{pb}$. In figure
\ref{fig:pipi1500} we show that the signal still has very strong significance for the heavy mass case, for $10~\rm{fb}^{-1}$ of integrated luminosity we
get $\chi_{sig}\sim 45$. We reiterate at this point that there is strong discovery potential for hyperpions heavier than $450~\rm{GeV}$ as well, however
in the benchmark model beyond this point the coloron pair production cross section is small and the analysis described in the next section becomes
unrealistic.

\subsection{Eight Jet Analysis}
\label{subsec:8j}
\FIGURE[!t]{\includegraphics[width=5.0in]{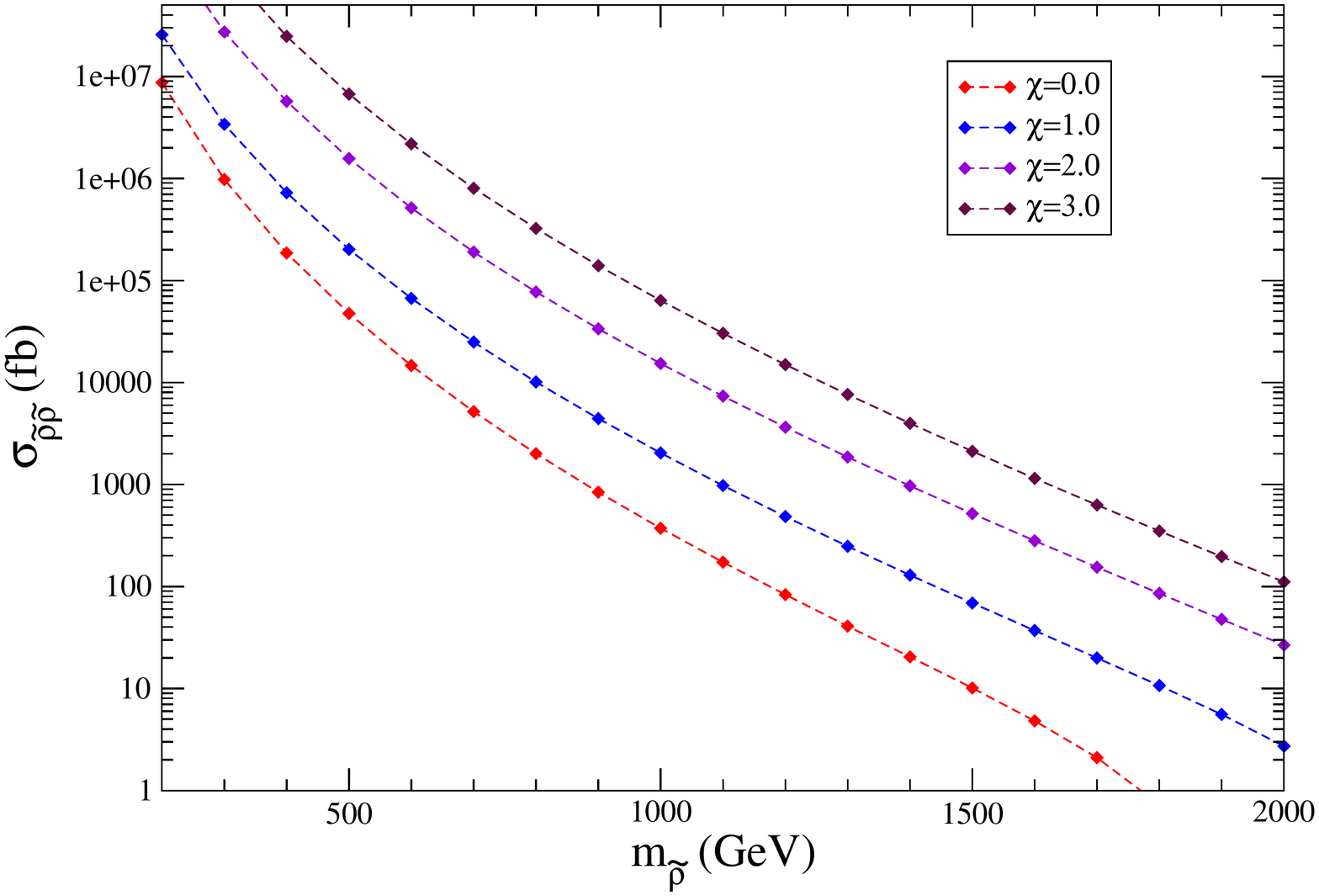} \caption{Coloron pair production cross section as a function of $m_{\col}$.
The various curves represent the impact of different choices of the parameter $\chi$.} \label{fig:coloronpairxsec}}
Even though we were able to demonstrate strong discovery potential for the hyperpions, we have not yet conclusively presented evidence for the existence
of the coloron itself. For this we search for coloron pair production. The pair production cross section is plotted in figure \ref{fig:coloronpairxsec}.
Note that the parameter $\chi$ contributes to both the three-point as well as four-point couplings of the coloron to gluons, thus the coloron pair
production amplitude has a piece proportional to $\chi^{2}$ and the cross section can grow as $\chi^{4}$. As discussed in section \ref{sec:model}, we adopt the choice $\chi=1$ as it appears to be the most plausible
one in regard of similar theories containing color octet spin-1 particles. We can see that the cross section is quite large which allows us to compete
with QCD backgrounds in terms of total cross section even before we impose kinematic cuts aimed at reconstructing the kinematic information in the signal
events.

We will present our analysis of the eight jet channel relying only on parton level cross section information for background, using a full 8-parton final
state integration obtained with \Comix. Generating a statistically large enough background sample of \textit{unweighted} 8-parton events is unfortunately
beyond our computational means, we will however confirm the validity of our conclusions in appendix \ref{app:8j-bg} using inclusive 6-parton events that
were then showered using Pythia, keeping in mind of course that kinematic shapes in the background are expected to have significant systematic
uncertainties.

\subsubsection{Lighter mass case}
\label{subsec:8j-light}

In order to decrease the background cross sections as much as possible we use an incremental $p_{T}$ cut scheme. We determine the cut values from a Monte
Carlo study of signal events, we choose a $p_{T}$ cut for the eight hardest jets such that the efficiency of the cut on the eighth hardest jet is
$\sim55\%$ and each additional cut has a relative efficiency of $\sim80\%$, this sets the efficiency of the combined cut scheme on the signal to be
$\sim13\%$. In particular for $m_{\col}=750~\rm{GeV}$ we demand $p_{T,j8}>40~\rm{GeV}$, $p_{T,j7}>60~\rm{GeV}$, $p_{T,j6}>90~\rm{GeV}$,
$p_{T,j5}>125~\rm{GeV}$, $p_{T,j4}>160~\rm{GeV}$, $p_{T,j3}>200~\rm{GeV}$, $p_{T,j2}>250~\rm{GeV}$, $p_{T,j1}>320~\rm{GeV}$. With these cuts the
background cross section is $1.2~\rm{pb}$ at parton level, and is expected to decrease further once the cuts are imposed at jet level, since parton
showers will typically split the jets and therefore reduce the $p_{T}$ of individual jets (this expectation will be confirmed in appendix
\ref{app:8j-bg}). The signal cross section after imposing the $p_{T}$ cuts past parton showering and jet reconstruction is $2.6~\rm{pb}$.

Despite the systematic uncertainties in computing the eight jet backgrounds, these numbers indicate that even without taking advantage of any kinematic
information in the events such as invariant masses of pairs of jets, signal and background cross sections are expected to be comparable in size even when
one allows for relatively large $K$-factors. Below we present a signal only study to demonstrate that with the proper analysis cuts we can reconstruct the
cascade decays in the signal events, while the efficiency for background events to pass these reconstruction cuts is expected to be very low. This will be
confirmed for showered 6-parton events in appendix \ref{app:8j-bg}.

First we do a Monte Carlo study of signal events in order to compare the parton level truth information to the result of parton showering, hadronization
and detector effects. We study the fate of the eight jets coming from the decay of the coloron pair and we find that fairly often parton showering will
split the jets and make the original kinematics very hard to reconstruct. Furthermore, with the high number of jets present in the final state it is not
uncommon for at least two of the primary jets to partially overlap and be reconstructed as a single jet. We do find however, that in the events where the
eight primary gluons are faithfully found as eight separate jets, they are virtually always among the ten hardest jets in the event, and the hardest four
jets almost never come from additional radiated partons. Based on this information, we design our analysis procedure as follows: We take the ten hardest
jets in an event (if there are only eight or nine jets in the event, we take those, while the rest of the procedure is unchanged), we then take all
possible subsets of eight jets out of those always keeping the hardest four, and we pair them up into four pairs in all possible ways. We then demand that
all four pairs have an invariant mass in a window that corresponds to the signal peak in the four jet analysis (the histogram in figure
\ref{fig:pipi750}). In the lighter coloron case this is $175~\rm{GeV}<m_{2j}<245~\rm{GeV}$. We drop events for which no such pairing exists. We expect
this cut to severely reduce background as background events normally display a large hierarchy of $p_{T}$'s and it is very unlikely that four pairs with
similar masses can be formed. In fact, considering the small pairing efficiency of background in the four jet analysis where we demand only two pairs with
similar invariant masses, we are confident that requiring four pairs with comparable masses will yield a virtually pure signal sample. For the background
analysis on 6-parton events presented in appendix \ref{app:8j-bg} this is shown to be valid.

\begin{table}
\begin{center}
\begin{tabular}{c||c|c|c|c}
$\sigma$ with & 1 pairing & 2 pairings & 3 pairings & more pairings\\
\hline (in pb) & 0.28 & 0.10 & 0.04 & 0.05
\end{tabular}
\end{center}
\caption{\label{tab:pairing-dist-750-signal}Cross section (after $p_{T}$ cuts) of signal events with one or multiple candidates for four dijet-pairs all
compatible with $175~\rm{GeV}<m_{2j}<245~\rm{GeV}$.}
\end{table}
The signal cross section after the invariant mass cuts is found to be $0.47~\rm{pb}$. While most events that pass the cuts have a unique pairing that
satisfies the cuts, there are events with more than one possible pairing, the distribution is given in table \ref{tab:pairing-dist-750-signal}. Instead of
throwing away the events with more than one possible pairing we choose to assign a weight to all possible pairings in an event that pass the invariant
mass cuts (i.e. where all four pairs fall in the correct mass window) and take a weighted average for the coloron masses as follows: For any pairing
$\mathcal{P}$ that passes the invariant mass cuts (as an example $\mathcal{P}$ could correspond to jets (1,5),(2,9),(3,4),(6,8) being paired up), let
$\lbrace m_{\ourpi,\mathcal{P},i}\rbrace$ be the values of the four reconstructed hyperpion masses. Now as we pair the hyperpions themselves, we get three
possible pairs of coloron masses, let these be $(m_{\col_{<},(\mathcal{P},j)}\,,m_{\col_{>},(\mathcal{P},j)})$, where $j=1,2,3$ (continuing with the above
example, the pair (1,5) could be paired with (2,9),(3,4) or (6,8)). We define the (unnormalized) weight for any such pairing as
\begin{equation}
w_{(\mathcal{P},j)}=\frac{1}{\sigma_{\ourpi,\mathcal{P}}\,\sigma_{\col,(\mathcal{P},j)}}\,,
\end{equation}
where
\begin{eqnarray}
\sigma_{\ourpi,\mathcal{P}}&=&\frac{1}{\langle m_{\ourpi,\mathcal{P}}\rangle^{2}}\sum_{i=1}^{4}\left(m_{\ourpi,\mathcal{P},i}-\langle m_{\ourpi,\mathcal{P}}\rangle\right)^{2}\nn\,,\\
\sigma_{\col,(\mathcal{P},j)}&=&\frac{1}{\langle m_{\col,(\mathcal{P},j)}\rangle^{2}}\sum_{i=1}^{2}\left(m_{\col,(\mathcal{P},j),i}-\langle
m_{\col,(\mathcal{P},j)}\rangle\right)^{2}.
\end{eqnarray}
\FIGURE[!t]{\includegraphics[width=5.0in]{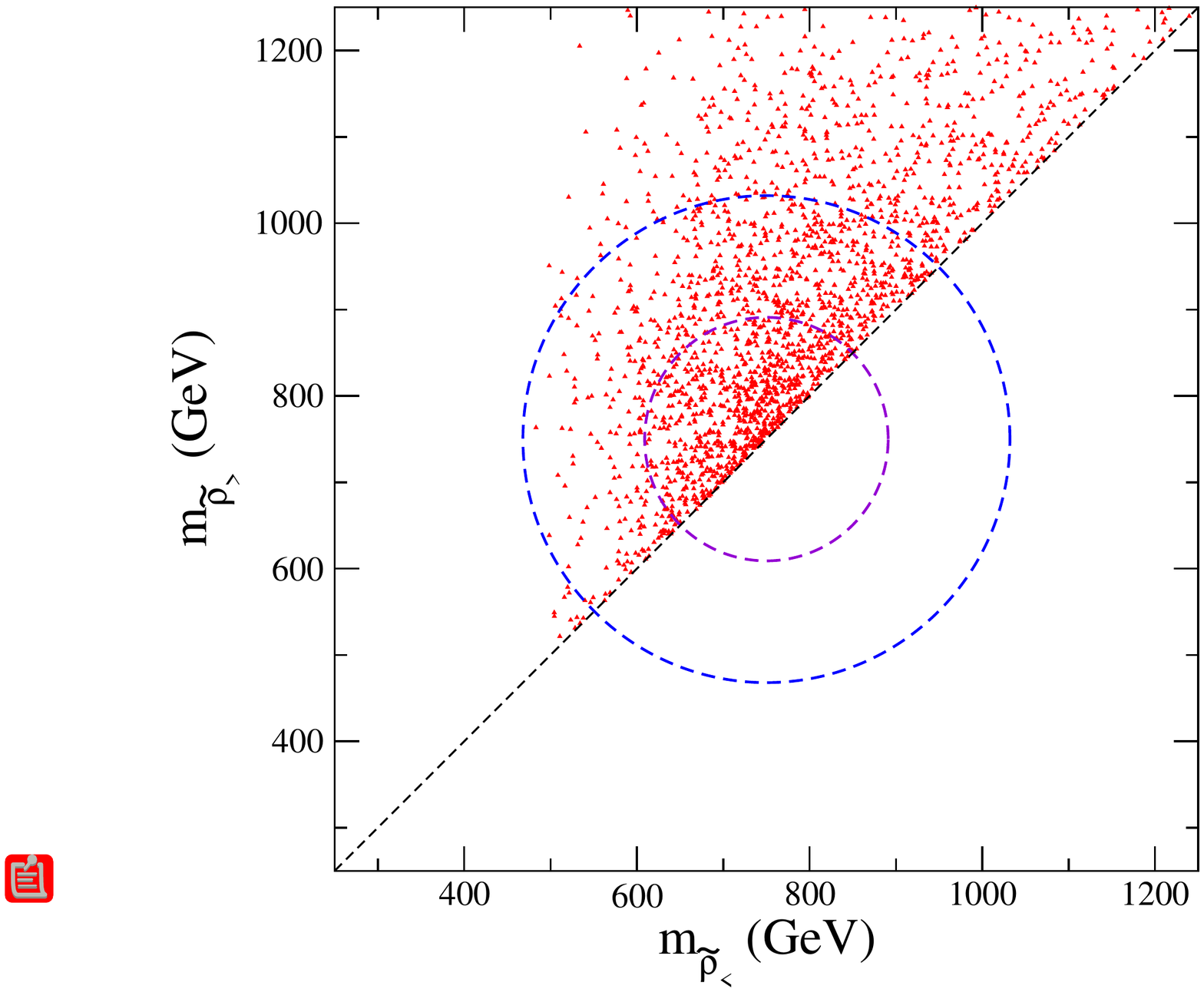} \caption{Coloron pair masses calculated from signal events with ($5~\rm{fb}^{-1}$) of
luminosity accumulate near the true mass. The two circles represent $1\Gamma_{\tilde\rho}$ and $2\Gamma_{\tilde\rho}$ distances from the true mass.}
\label{fig:coloronpairmass750}}
This weighting scheme favors pairings for which the hyperpion masses come out close to each other, and similarly it favors the pairing of the hyperpions
which yields coloron masses that are close to each other.

For each event we then calculate one set of coloron masses as the weighted average
\begin{eqnarray}%
\langle m_{\col_{<}}\rangle\equiv\frac{\sum_{(\mathcal{P},j)}\ w_{(\mathcal{P},j)}\,m_{\col_{<},(\mathcal{P},j)}}{\sum_{(\mathcal{P},j)}\ w_{(\mathcal{P},j)}}\nn\,,\\
\langle m_{\col_{>}}\rangle\equiv\frac{\sum_{(\mathcal{P},j)}\ w_{(\mathcal{P},j)}\,m_{\col_{>},(\mathcal{P},j)}}{\sum_{(\mathcal{P},j)}\
w_{(\mathcal{P},j)}}\,.
\end{eqnarray}
The results are plotted in figure \ref{fig:coloronpairmass750} for $5~\rm{fb}^{-1}$ of data. The bulk of the signal is indeed reconstructed near the true
mass for the coloron. Note that the main input for making this plot is the extraction of the hyperpion mass from the 4-jet analysis, a feature of the
\textit{signal}. Thus there is no reason to expect the background to have an accumulation point, the showered 6-parton background events in appendix
\ref{app:8j-bg} confirm this expectation.

\subsubsection{Heavier mass case}
\label{subsec:8j-heavy}

\begin{table}
\begin{center}
\begin{tabular}{c||c|c|c|c}
$\sigma$ with & 1 pairing & 2 pairings & 3 pairings & more pairings\\
\hline (in fb) & 1.1 & 0.32 & 0.14 & 0.17
\end{tabular}
\end{center}
\caption{\label{tab:pairing-dist-1500-signal}Cross section (after $p_{T}$ cuts) of signal events with one or multiple candidates for four dijet-pairs all
compatible with $350~\rm{GeV}<m_{2j}<475~\rm{GeV}$.}
\end{table}
We use the same techniques for the case of the heavier coloron. Even though we are pair producing $1.5~\rm{TeV}$ colorons, we find that the signal cross
section is large enough for this analysis to be possible. The $p_{T}$ cut scheme we are using for the heavier mass coloron is given by
$p_{T,j8}>75~\rm{GeV}$, $p_{T,j7}>115~\rm{GeV}$, $p_{T,j6}>180~\rm{GeV}$, $p_{T,j5}>240~\rm{GeV}$, $p_{T,j4}>300~\rm{GeV}$, $p_{T,j3}>380~\rm{GeV}$,
$p_{T,j2}>470~\rm{GeV}$, $p_{T,j1}>600~\rm{GeV}$. With these cuts, the background cross section at parton level is $5.7~\rm{fb}$, which once again is
expected to drop further once the cuts are imposed after parton showering and jet reconstruction. The signal cross section after showering and jet
reconstruction is $9.5~\rm{fb}$.

\FIGURE[!t]{\includegraphics[width=5.0in]{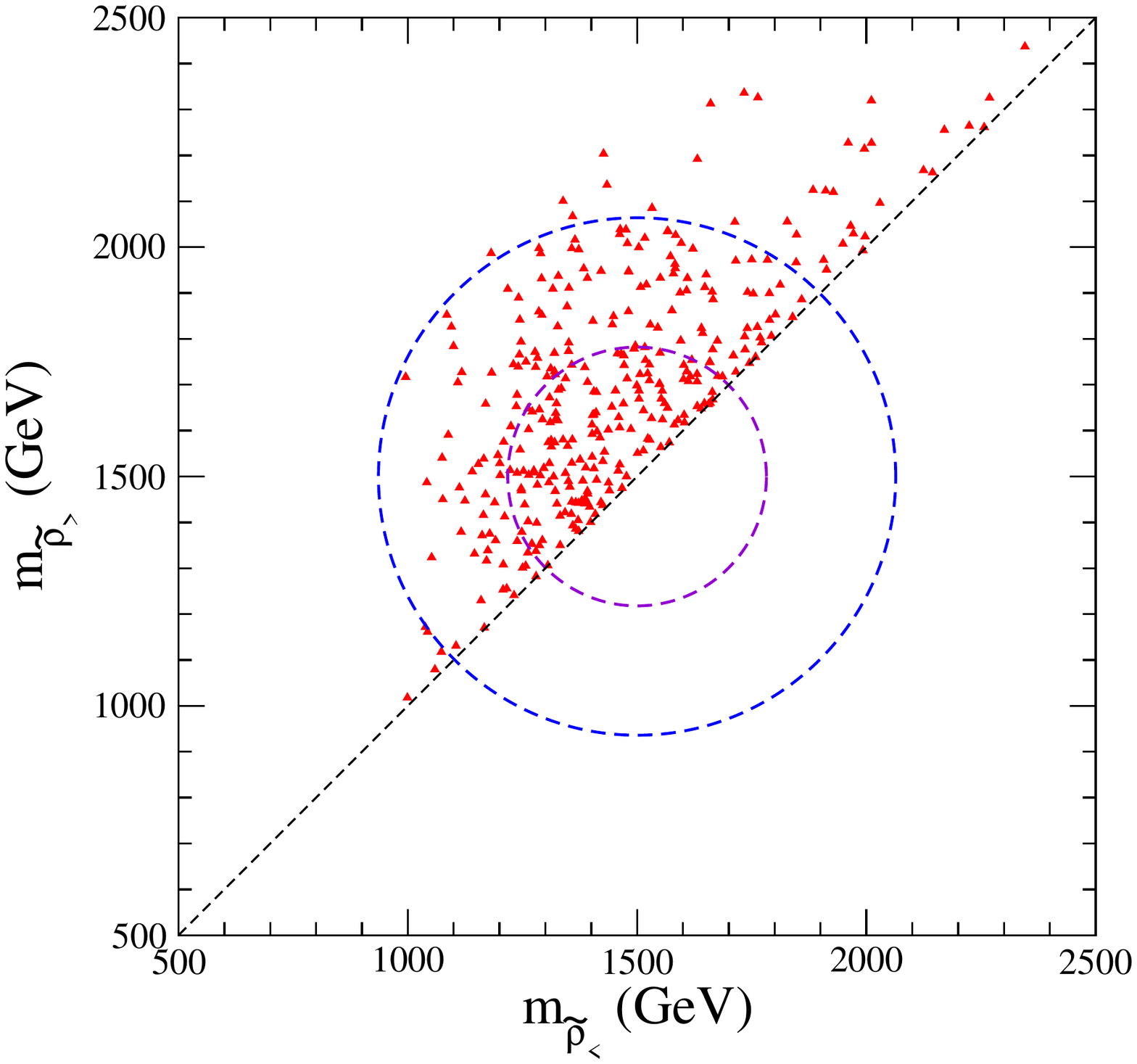} \caption{Coloron pair masses calculated from signal events ($200~\rm{fb}^{-1}$) accumulate
near the true mass. The two circles represent $1\Gamma_{\tilde\rho}$ and $2\Gamma_{\tilde\rho}$ distances from the true mass.}
\label{fig:coloronpairmass1500}}
We then use the mass window $350~\rm{GeV}<m_{2j}<475~\rm{GeV}$ (in accordance with figure \ref{fig:pipi1500}) to veto events in which the jets do not pair
up correctly, as described for the lighter mass case. Once again, the majority of the events past the invariant mass cuts have a unique pairing, as
demonstrated in table \ref{tab:pairing-dist-1500-signal}. Using the same technique as in the lighter mass case, we find that with $200~\rm{fb}^{-1}$ of
data one can reconstruct signal events quite successfully. Our results are plotted in figure \ref{fig:coloronpairmass1500}.

\subsection{Guidelines for a Self Calibrating Search}
\label{subsec:generic search} While in sections \ref{subsec:4j} and \ref{subsec:8j} we have highlighted the discovery potential of the LHC for the
hyperpions and colorons for a wide range of masses, the $p_{T}$ cuts in each case were chosen to enhance signal over background. In the following, we will
attempt to prescribe a search strategy that can be used in an experimental search without prior knowledge of the scale of new physics. We will also
address which stages of the analysis could be susceptible to large systematic errors in our background estimates relying on tree level amplitudes.

The only input entering the four jet analysis we presented in section \ref{subsec:4j} was the $p_{T}$ cuts on the four hardest jets. (We find that
demanding the two hyperpion candidate masses to be within $50~\rm{GeV}$ of each other works for the entire mass range of interest.) However, the
prominence of the hyperpion mass peak above background in figures \ref{fig:pipi750} and \ref{fig:pipi1500} demonstrates that this analysis can be done
with a sliding $p_{T}$ cut. As the cut value is increased, the background distribution will shift to higher energies and the signal peak separates cleanly
from the background for reasons described in section \ref{subsec:4j}. Thus the four jet analysis can be modularized, and the position of the hyperpion
mass peak can be used as an input to the eight jet analysis, where we demand four pairs of jets with an invariant mass in the window determined by the
four jet analysis. Note also that this procedure is robust to errors of $\mathcal{O}(10)$ in the background estimates, since $\chi_{sig}$ scales as
$B^{-1/2}$, therefore even with a $K$-factor of 10 the statistical significance in both our benchmark studies remains above $10\sigma$.

\FIGURE[!t]{
\begin{picture}(500,90)
\put(300,-20){\includegraphics[width=5.8cm]{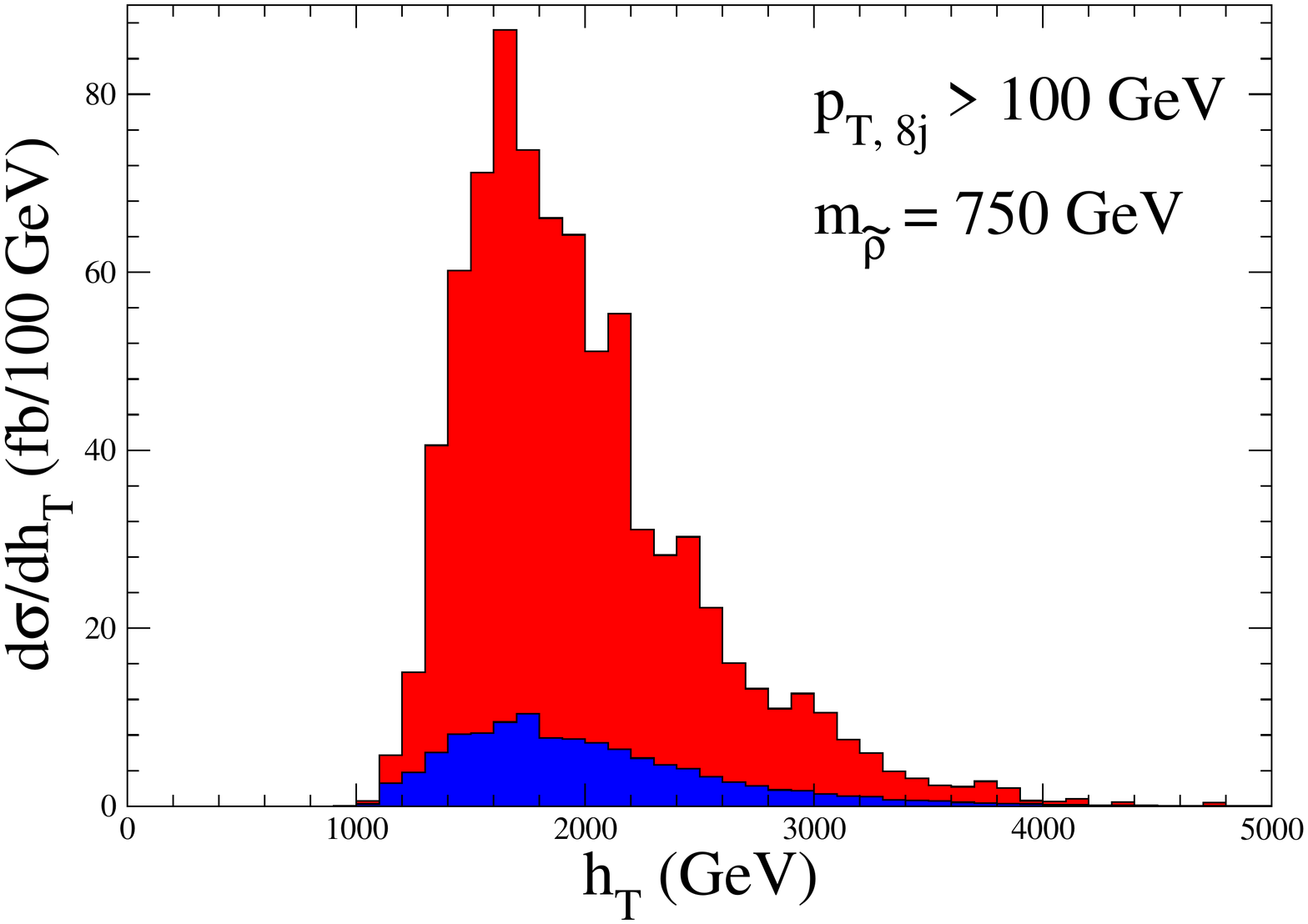}} 
\put(150,-20){\includegraphics[width=5.8cm]{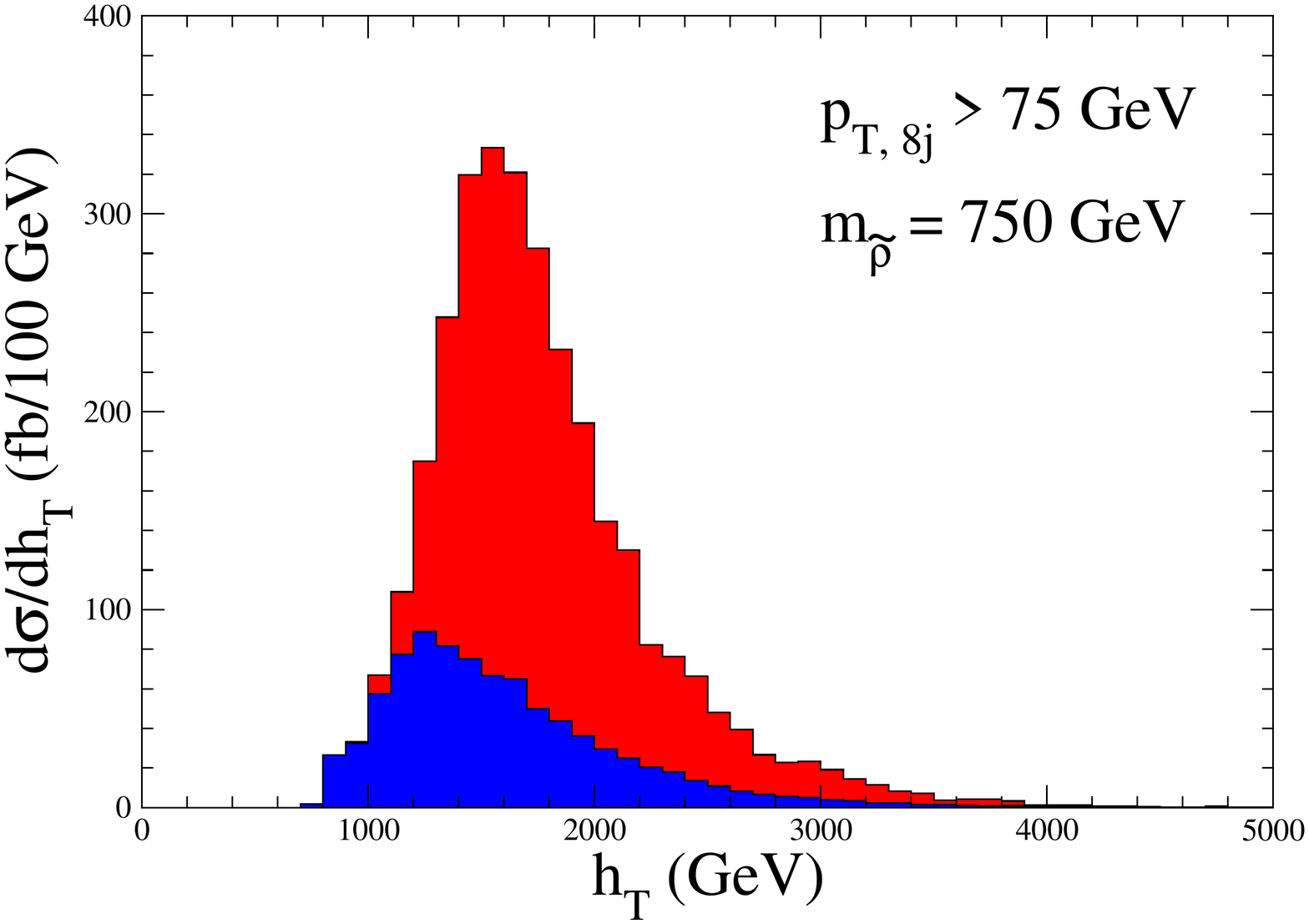}} 
\put(0,-20){\includegraphics[width=5.8cm]{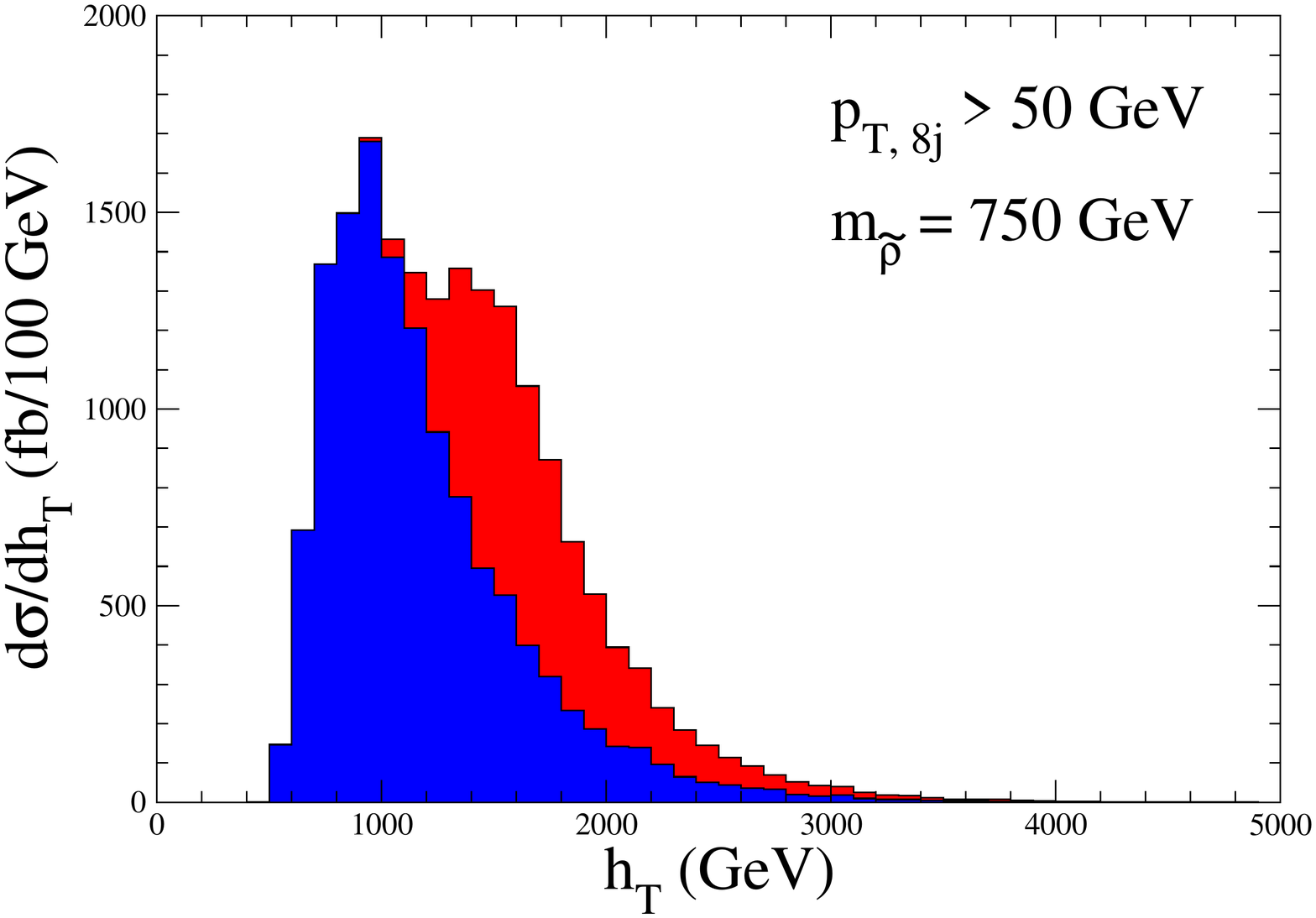}}
\end{picture}
\caption{\label{fig:htsequence750}The $h_{T}$ distribution of signal (red) and background (blue) with a cut on the $p_{T}$ of the eighth hardest jet.
Signal cross sections represent coloron pair production with $m_{\col}=750~\rm{GeV}$. From left to right, the value of the $p_{T}$ cut is increased from
50~GeV to 75~GeV to 100~GeV.} }
\FIGURE[!t]{
\begin{picture}(500,90)
\put(300,-20){\includegraphics[width=5.8cm]{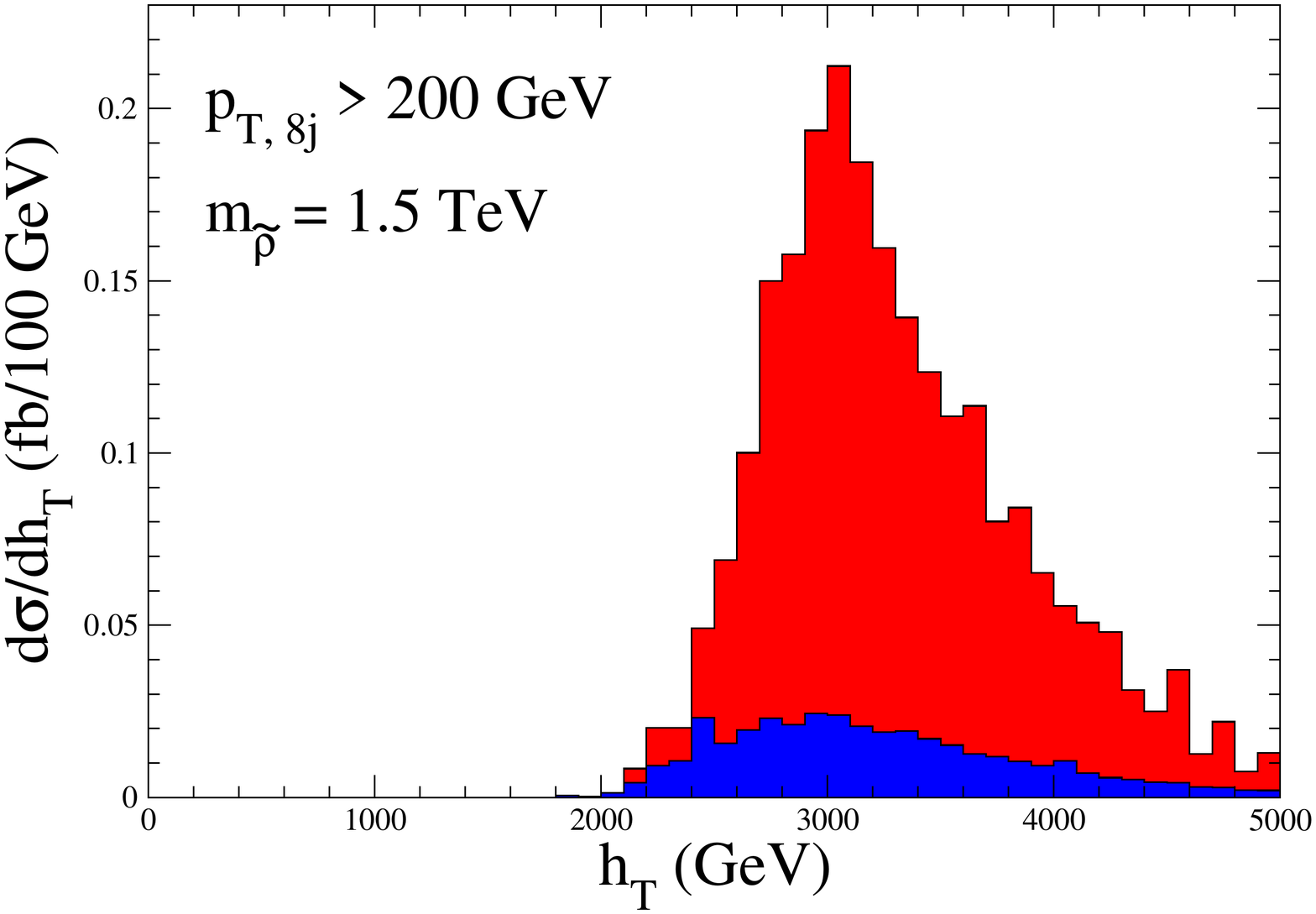}} 
\put(150,-20){\includegraphics[width=5.8cm]{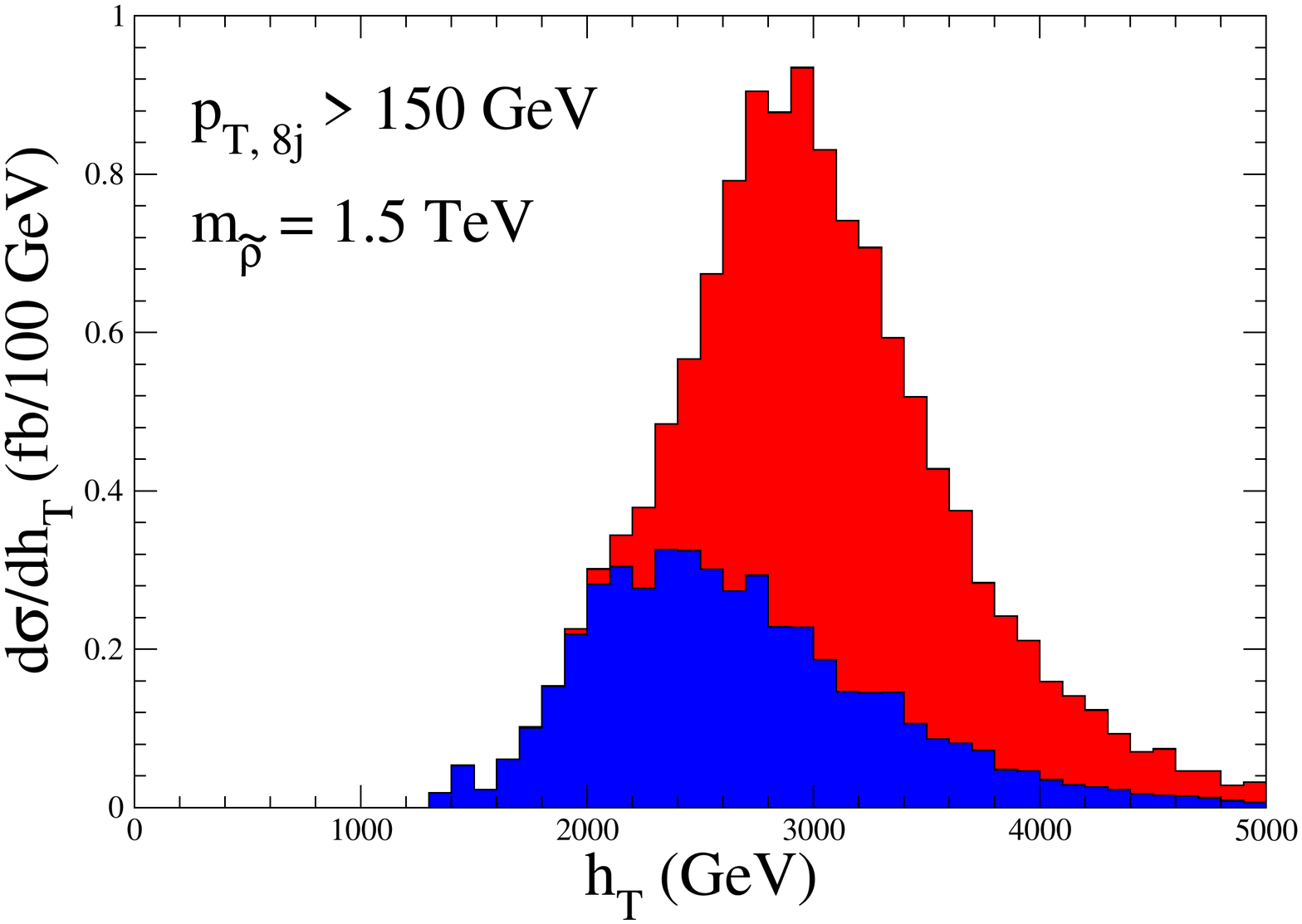}} 
\put(0,-20){\includegraphics[width=5.8cm]{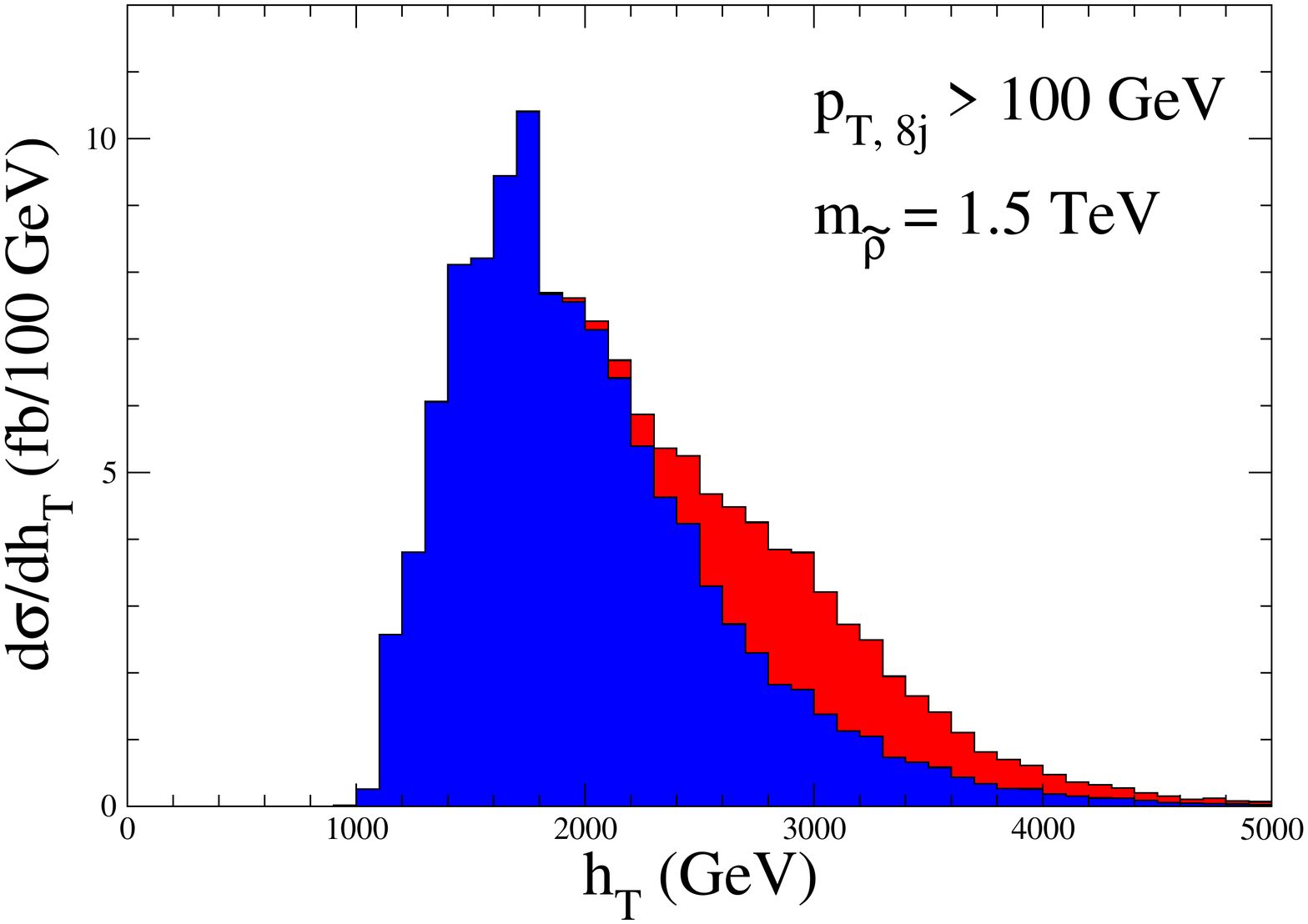}}
\end{picture}
\caption{\label{fig:htsequence1500}The $h_{T}$ distribution of signal (red) and background (blue) with a cut on the $p_{T}$ of the eighth hardest jet.
Signal cross sections represent coloron pair production with $m_{\col}=1.5~\rm{TeV}$. From left to right, the value of the $p_{T}$ cut is increased from
100~GeV to 150~GeV to 200~GeV.} }

Note that this way of performing the four jet analysis is very general and does not assume the existence of the coloron. Since the signal cross section is
dominated by the contribution from minimal QCD couplings of a color octet scalar, we have demonstrated strong discovery potential for any state which is
pair produced with a comparable cross section and which dominantly decays to dijets, with the implicit assumption that the coupling giving rise to this
decay mode is small, otherwise this state may first be observed as a dijet resonance. A similar analysis for states appearing in non-minimal technicolor
models has been performed in ref.~\cite{Chivukula:1991zk}. Spin-1 states decaying to dijets and scalars decaying to heavy flavors have been looked at in
ref.~\cite{Dobrescu:2007yp}. Even though ref.~\cite{Zerwekh:2008mn} investigated pair production of pseudoscalars that predominantly decay to four jets,
they concentrated their efforts on a subdominant decay channel including a photon. The phenomenology of ref.~\cite{Plehn:2008ae} also includes a resonance
channel with a four jet final state, however they focus in their analysis on final states with top quarks.

\FIGURE[!t]{\includegraphics[width=5.0in]{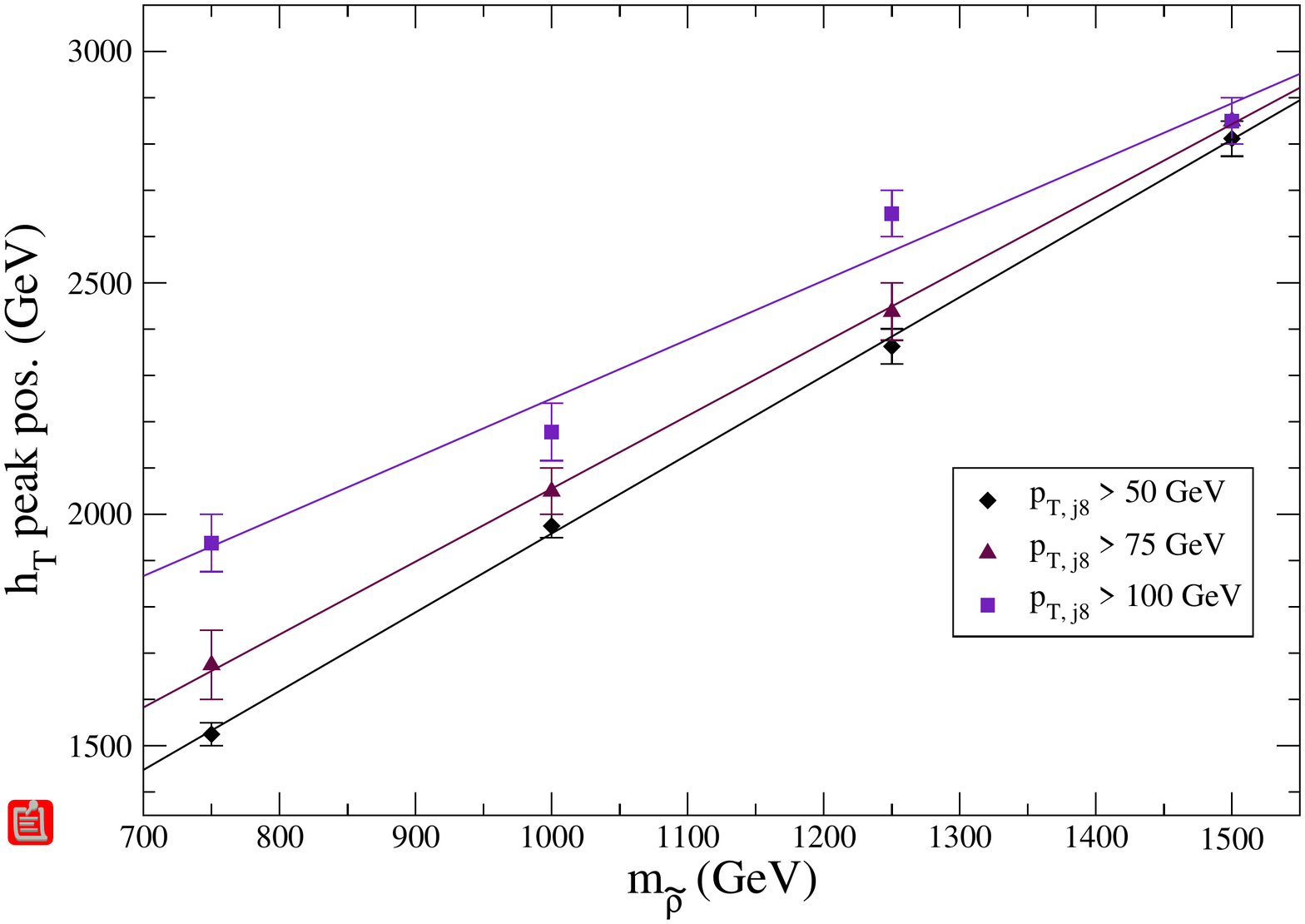} \caption{The peak position of the $h_{T}$ distribution in the signal as a function of the
coloron mass, for different values of the $p_{T}$ cut on the eight hardest jet. The behavior is linear to a good approximation. The error bars indicate
the uncertainty in specifying a precise position for where the distribution peaks, and become larger as the number of events in the distribution decrease
as a consequence of the increasing $p_{T}$ cut.} \label{fig:htpeakvsmcol}}
To demonstrate how the optimized $p_{T}$ cuts entering the eight jet analysis can be obtained, we focus on the $h_{T}$ distribution of eight jet events
with a sliding $p_{T}$ cut on the eighth hardest jet. ($h_{T}$ is defined as the scalar sum of all jet transverse momenta.) The results are displayed for
the lighter and heavier coloron cases in figures \ref{fig:htsequence750} and \ref{fig:htsequence1500}. At a particular value of the $p_{T}$ cut we find
that signal becomes visible as a bump above background (left plot) with a peak that is shifted away from the artificial peak in the background introduced
by the cuts. As the $p_{T}$ cut is increased further, signal falls less sharply than background (middle plot), and while at large values of the cut signal
overwhelms background in cross section, the difference between signal and background becomes one of normalization rather than one of shape (right plot)
since the $p_{T}$ cut begins to shift the signal peak as well. While these plots are made at parton level, we expect parton showering to affect signal and
background shapes similarly since there should be no significant difference in the way they radiate, and our conclusions should still apply. The point we
wish to make here is that even if the background estimates needed to be scaled up by a $K$-factor of $\mathcal{O}(1)$, the signal would still be visible
as a bump at a higher value of the $p_{T}$ cut (e.g. in the middle plot rather than the left plot) due to the different rate at which the signal and
background cross sections fall with the increasing cut value. This procedure can break down if the $K$-factor is so large ($\mathcal{O}(10)$) that the
signal cross section would only become comparable to background by the time that the $p_{T}$ cut is large enough to shift the signal peak position to
match the background shape (right plot), in which case there may be no clear indication of a signal bump.

Assuming that the systematic errors in our background estimates are not so large as to invalidate the procedure described above, the position of the
$h_{T}$ peak can be used to deduce the coloron mass, as they are linearly correlated. This is displayed in figure \ref{fig:htpeakvsmcol} where we plot the
peak position of the $h_{T}$ distribution in the signal as a function of the coloron mass using events that have been passed through \Pythia and \PGS.
While this plot has a dependence on the jet reconstruction algorithm that is used as well as details of detector performance, it demonstrates that a
linear relation exists between the signal peak of the $h_{T}$ distribution and the mass of the coloron. We expect that any experimental search would
reproduce figure \ref{fig:htpeakvsmcol} using their own jet reconstruction algorithm and detector simulation.

\FIGURE[!t]{\includegraphics[width=5.0in]{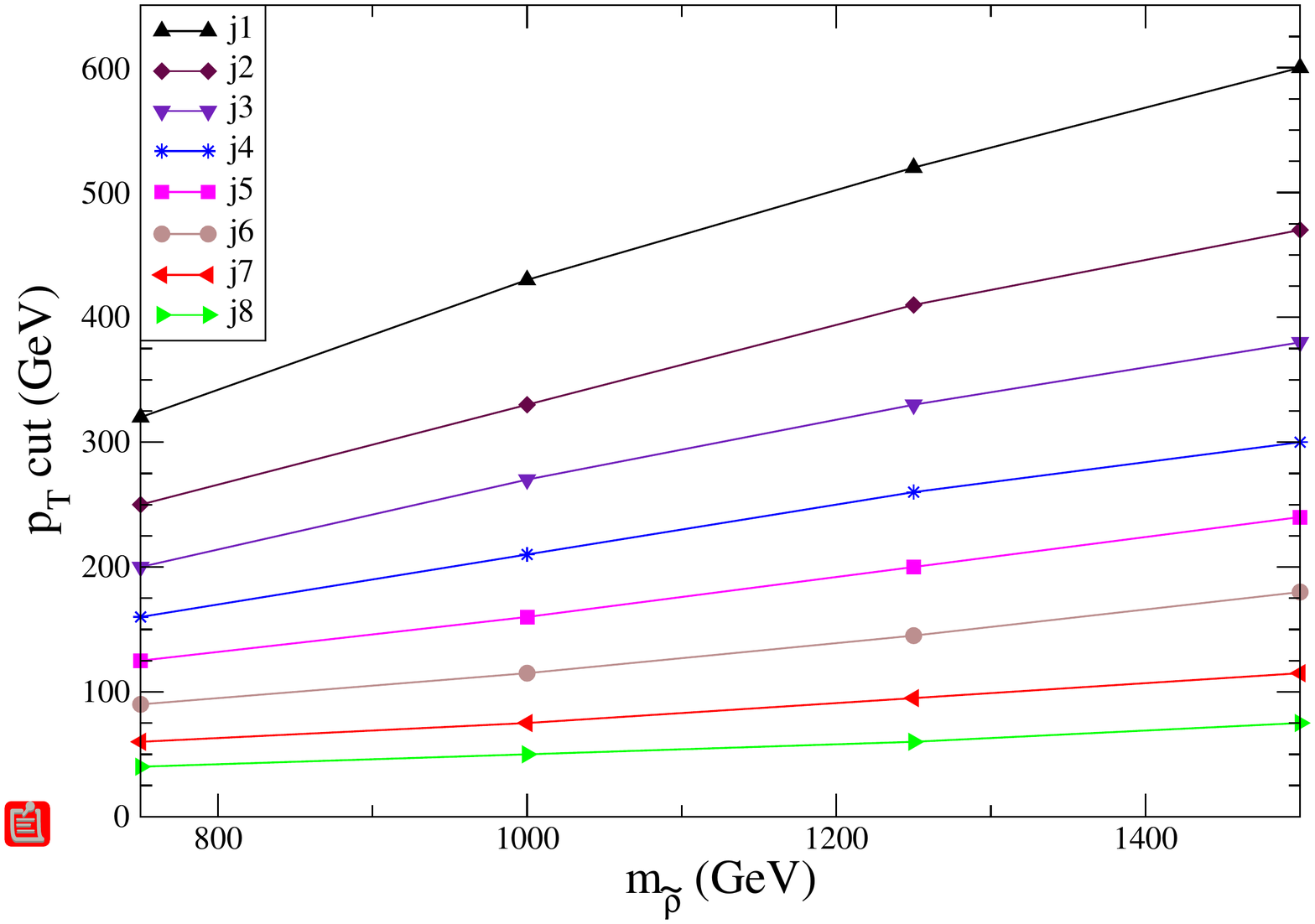} \caption{A scheme for picking the incremental $p_{T}$ cuts on the eight hardest jets as
a function of the coloron mass. The values are chosen such that the efficiency of the cut on the eighth hardest jet on signal events is 55\%, and the
efficiency of each consecutive cut is 80\%, such that the total efficiency of this cut scheme on signal is 12\%. We find that this choice gives a higher
S/B enhancement compared to a single $p_{T}$ cut on the eight hardest jets.} \label{fig:staggeredcutsvsmcol}}
This procedure makes it possible to link the excess in the $h_{T}$ distribution to a candidate mass for the coloron and use that to optimize the
incremental $p_{T}$ cuts that enter the eight jet analysis. Note that our analysis does not assume $\frac{m_{\ourpi}}{m_{\col}}=0.3$ and is therefore
applicable to more general scenarios than the benchmark model. In figure \ref{fig:staggeredcutsvsmcol} we display a scheme for choosing the incremental
$p_{T}$ cuts that enter the eight jet analysis, the values of the $p_{T}$ cuts are chosen for each value of $m_{\col}$ to have the same efficiency on the
signal as described in section \ref{subsec:8j}. Combining this with the result of the four jet analysis to set the pairing mass window, the eight jet
analysis can now be performed.

Note that while we have proposed using the $h_{T}$ distribution to determine the $p_{T}$ cuts to be used in the coloron mass reconstruction analysis, this
is not an essential step. Even with large errors on the background, we have shown that the eight jet signal and background cross sections are comparable
before any reconstruction cuts are imposed, which are expected to further enhance signal over background significantly, this is confirmed in appendix
\ref{app:8j-bg} with a background sample of showered 6-parton events. With this in mind, it should be possible to perform the coloron reconstruction
analysis with a sliding $p_{T}$ cut on the eight hardest jets instead of our optimized incremental $p_{T}$ cut scheme. While this may decrease $S/B$ by an
$\mathcal{O}(1)$ factor, the overall success of the analysis presented in section \ref{subsec:8j} should be robust.

\section{Conclusions}
\label{sec:conclusions}

Even though the production of colored states is the main strength of hadron colliders, at the present there exist no dedicated searches for new physics in
multijet channels because of the large background cross sections and the presence of systematic uncertainties in background estimates based on tree level
calculations. Continuing in the spirit of ref.~\cite{Kilic:2008pm}, we have reiterated in this paper that there is a family of theoretically very simple
models, relying on a minimal set of assumptions on how the SM is extended, with a subtle phenomenology where the states of new physics appear almost
exclusively in multijet channels. While these models can incorporate diverse choices for the new physics sector, the existence of colored octet vectors
and scalars as bound states of the underlying dynamics is generic.

We used a phenomenological Lagrangian to describe the production and decay of these states. There is a choice for the microscopic physics for which the
strong dynamics are an exact copy of QCD which allows us to extract the effective parameters of the phenomenological Lagrangian. We utilized this
benchmark model to show quantitatively that the collider phenomenology is interesting and argued that this conclusion remains generic for variations
around this choice. We then demonstrated for two choices of the coloron mass that hyperpions can be discovered in the four jet channel with relatively low
integrated luminosity. We have furthermore shown that colorons can be discovered in the eight jet channel with additional statistics. While we lack the
computational power to simulate a statistically large enough sample of unweighted 8-parton background events, we did demonstrate for a sample of 6-parton
events with subsequent showering that our expectations are confirmed.

We have further argued that the signal can be self calibrating and we have proposed a realistic search strategy that assumes no prior knowledge of the
scale of new physics, where the hyperpion search can be performed with a sliding $p_{T}$ cut, and an excess in the $h_{T}$ distribution of eight jet
events can be used to optimize the $p_{T}$ cuts for coloron mass reconstruction. While this last step may be susceptible to large errors in our background
estimates, it is not essential for the discovery of the coloron.

We remark that the collider phenomenology studied here can also be used in a more agnostic context. One of the most obvious new physics scenarios to
expect at a hadron collider is the pair production of a heavy colored state. Barring the case where this state is quasi-stable (which gives rise to
R-hadron signatures) it can decay in a number of ways, and since the decaying particle is colored, the decay into dijets is one of the most robust
possibilities. Once again, with this in mind it is surprising that model-independent collider searches for new physics in multijets have not been
performed in the past. We would like to emphasize that the four jet analysis we presented is independent from the existence of the coloron, thus we have
in fact demonstrated the discovery potential for the pair production of any colored state which is produced with a cross section comparable with a color
octet scalar and which decays to dijets, with the assumption that the coupling giving rise to this decay mode is small, otherwise the primary production
mechanism would be resonant production, and this state would have first manifested itself as a dijet resonance.

The LHC-period is a crucial time in particle physics, and while we may expect to discover a simple mechanism that solves the hierarchy problem, we should
be prepared for discoveries in other channels which may (at first) appear unrelated to such a mechanism. The models presented here are simple, they can
occur readily in gauge field theory, they are consistent with all experimental constraints and their phenomenology overlaps with the main strength of the
LHC, the production of colored new particles. Whether or not they may be connected to a larger mechanism having to do with the electroweak symmetry
breaking, they provide a template for a new physics scenario appearing in the multijet channel. In this work we have demonstrated that the LHC has strong
discovery potential for such new physics, even in the case that it may arise from a different underlying mechanism than the one we have used to model our
phenomenological Lagrangian on. Therefore we propose that multijet resonance searches, like dijet resonance searches, should become standard at hadron
colliders.

\begin{acknowledgments}
The authors are indebted to Raman Sundrum for invaluable discussions and advice. C.K. and M.S. would also like to thank Johan Alwall, Kenichi Hatakeyama,
David E.~Kaplan, Kyoungchul Kong, Fabio Maltoni, Kirill Melnikov, Takemichi Okui and Brock Tweedie for useful suggestions, insights and comments. C.K.
would further like to thank the Aspen Center for Physics for providing a stimulating environment where part of the research in this work was completed.
S.S. wants to thank T.~Gr\'egoire and T.~Plehn for valuable discussions and S.~H\"oche for support related to using \Comix. C.K. and M.S. are supported by
the National Science Foundation grant NSF-PHY-0401513 and the Johns Hopkins Theoretical Interdisciplinary Physics and Astrophysics Center. C.K. is further
supported in part by DOE grant DE-FG02-03ER4127 and by the Alfred P.~Sloan Foundation. The work of S.S. is supported by the UK Science and Technology
Facilities Council (STFC).
\end{acknowledgments}

\appendix
\section{Check of Our Conclusions Using Background Events}
\label{app:8j-bg} Even though we use \Comix for a full 8-parton phase space integration, generating a sample of unweighted 8-parton events is beyond our
computational means. In order to test the validity of our conclusions we have therefore chosen to generate a sample of 6-parton events which were
subsequently showered, hadronized and reconstructed using \Pythia and PGS. Since these results should only be taken as a sanity check rather than a fully realistic
background study, we choose to present them outside the main body of this paper.

\subsection{Lighter Mass Case}

\begin{table}
\begin{center}
\begin{tabular}{c||c|c|c|c}
$\sigma$ with & 1 pairing & 2 pairings & 3 pairings & more pairings\\
\hline (in fb) & 17 & 6.8 & 2.0 & 2.6
\end{tabular}
\end{center}
\caption{\label{tab:pairing-dist-750-bg}Cross section (after $p_{T}$ cuts) of background events with one or multiple candidates for four dijet-pairs all
compatible with $175~\rm{GeV}<m_{2j}<245~\rm{GeV}$.}
\end{table}
After the incremental $p_{T}$ cuts described in section \ref{subsec:8j-light}, we find that the background cross section to be $1.0~{\rm pb}$, lower than
the parton level result as expected. Upon demanding the presence of four pairs of jets in the mass window $175~\rm{GeV}<m_{2j}<245~\rm{GeV}$ we find that
$97\%$ of the background is eliminated, the cross section of background events with one or more possible pairing is given in table
\ref{tab:pairing-dist-750-bg}. Using the reconstruction techniques described in section \ref{subsec:8j-light} we superimpose the background events on top
of the signal in figure \ref{fig:coloronpair750withbg}. It is evident that background events do not have an accumulation point while the signal favors the
true coloron mass.
\FIGURE{\includegraphics[width=5.0in]{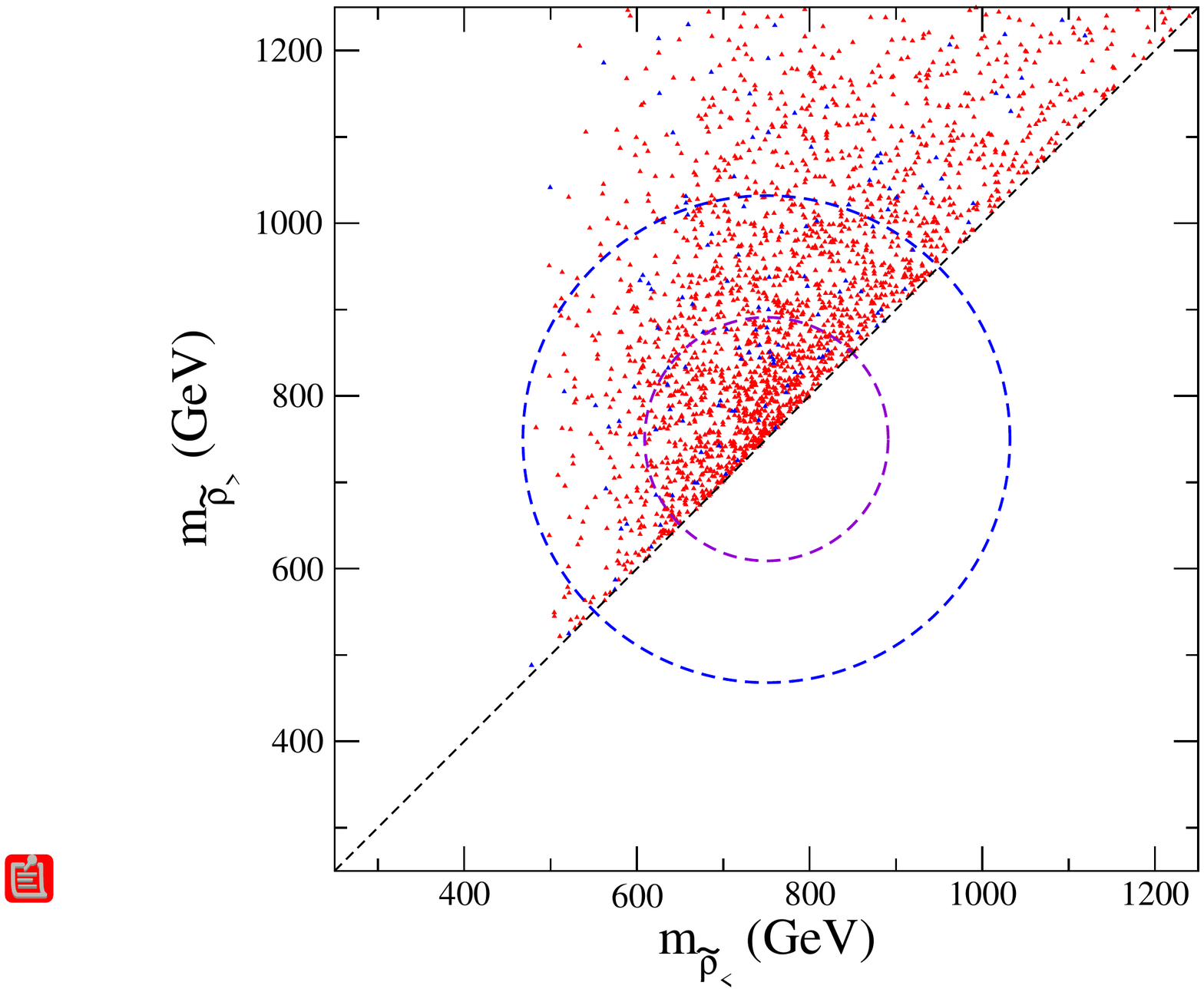} \caption{Coloron pair masses calculated from signal (red) and background (blue) events
with ($5~\rm{fb}^{-1}$) of luminosity. While signal accumulates near the true value of the coloron mass, there is no obvious accumulation point for the
background. The two circles represent $1\Gamma_{\tilde\rho}$ and $2\Gamma_{\tilde\rho}$ distances from the true mass.} \label{fig:coloronpair750withbg}}

\subsection{Heavier Mass Case}

\begin{table}
\begin{center}
\begin{tabular}{c||c|c|c|c}
$\sigma$ with & 1 pairing & 2 pairings & 3 pairings & more pairings\\
\hline (in fb) & 0.14 & 0.05 & 0.01 & 0.02
\end{tabular}
\caption{\label{tab:pairing-dist-1500-bg}Cross section (after $p_{T}$ cuts) of background events with one or multiple candidates for four dijet-pairs all
compatible with $350~\rm{GeV}<m_{2j}<475~\rm{GeV}$.}
\end{center}
\end{table}
Using the incremental $p_{T}$ cuts described of section \ref{subsec:8j-heavy}, the background cross section is found to be $4.9~{\rm fb}$, lower than the
parton level result in accordance with expectations. Requiring the presence of four pairs of jets in the mass window $350~\rm{GeV}<m_{2j}<475~\rm{GeV}$
eliminates $96\%$ of the background, the remaining cross section being distributed as in table \ref{tab:pairing-dist-1500-bg}. The results after
reconstructing the coloron mass are displayed in figure \ref{fig:coloronpair1500withbg}. Once again, signal accumulates near the true coloron mass while
the background has no obvious point of accumulation.
\FIGURE{\includegraphics[width=5.0in]{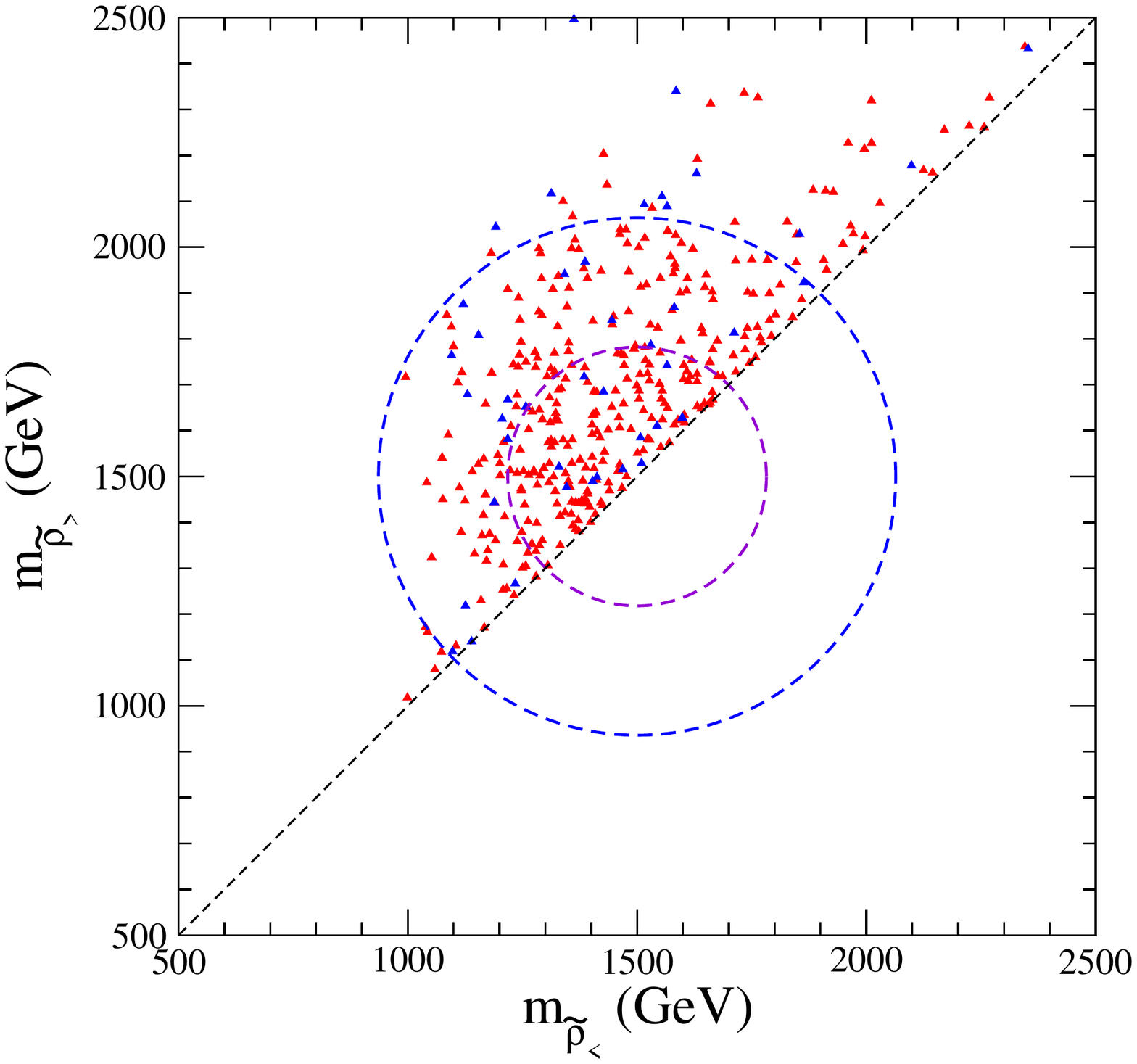} \caption{Coloron pair masses calculated from signal (red) and background (blue) events
($200~\rm{fb}^{-1}$). Signal accumulates at the true mass while there is no obvious accumulation point for background. The two circles represent
$1\Gamma_{\tilde\rho}$ and $2\Gamma_{\tilde\rho}$ distances from the true mass.} \label{fig:coloronpair1500withbg}}
%



\begin{thebibliography}{99}

\bibitem{Kilic:2008pm}
  C.~Kilic, T.~Okui and R.~Sundrum,
  JHEP {\bf 0807}, 038 (2008)
  [arXiv:0802.2568 [hep-ph]].

\bibitem{coloron-1}
  C.~T.~Hill,
  Phys.\ Lett.\  B {\bf 266}, 419 (1991).

\bibitem{coloron-2}
  R.~S.~Chivukula, A.~G.~Cohen and E.~H.~Simmons,
  Phys.\ Lett.\  B {\bf 380}, 92 (1996)
  [arXiv:hep-ph/9603311].

\bibitem{Farhi:1979zx}
  E.~Farhi and L.~Susskind,
  Phys.\ Rev.\  D {\bf 20}, 3404 (1979).

\bibitem{Eichten:1984eu}
  E.~Eichten, I.~Hinchliffe, K.~D.~Lane and C.~Quigg,
  Rev.\ Mod.\ Phys.\  {\bf 56}, 579 (1984)
  [Addendum-ibid.\  {\bf 58}, 1065 (1986)].

\bibitem{Lane:2002sm}
  K.~Lane and S.~Mrenna,
  Phys.\ Rev.\  D {\bf 67}, 115011 (2003)
  [arXiv:hep-ph/0210299].

\bibitem{Manohar:1983md}
  A.~Manohar and H.~Georgi,
  Nucl.\ Phys.\  B {\bf 234}, 189 (1984).

\bibitem{Georgi:1985kw}
  H.~Georgi,
{\it  Menlo Park, Usa: Benjamin/cummings ( 1984) 165p.}

\bibitem{DelDebbio:2008zf}
  L.~Del Debbio, A.~Patella and C.~Pica,
  arXiv:0805.2058 [hep-lat].

\bibitem{Abe:1997hm}
  F.~Abe {\it et al.}  [CDF Collaboration],
  Phys.\ Rev.\  D {\bf 55}, 5263 (1997)
  [arXiv:hep-ex/9702004].

\bibitem{Abazov:2003tj}
  V.~M.~Abazov {\it et al.}  [D0 Collaboration],
  Phys.\ Rev.\  D {\bf 69}, 111101 (2004)
  [arXiv:hep-ex/0308033].

\bibitem{Hatakeyama:2008tz}
  K.~Hatakeyama and f.~t.~C.~Collaboration,
  arXiv:0810.3681 [hep-ex].

\bibitem{Cardaci:2008vg}
  M.~Cardaci  [ATLAS Collaboration and CMS Collaboration],
  arXiv:0805.2906 [hep-ex].

\bibitem{Bhatti:2008hz}
  A.~Bhatti {\it et al.},
  arXiv:0807.4961 [hep-ex].

\bibitem{Frampton:1987dn}
  P.~H.~Frampton and S.~L.~Glashow,
  Phys.\ Lett.\  B {\bf 190}, 157 (1987).

\bibitem{Gleisberg:2003xi}
  T.~Gleisberg, S.~H\"oche, F.~Krauss, A.~Sch\"alicke, S.~Schumann and J.~C.~Winter,
  JHEP {\bf 0402}, 056 (2004)
  [arXiv:hep-ph/0311263].

\bibitem{Krauss:2001iv}
  F.~Krauss, R.~Kuhn and G.~Soff,
  JHEP {\bf 0202}, 044 (2002)
  [arXiv:hep-ph/0109036].

\bibitem{Pumplin:2002vw}
  J.~Pumplin, D.~R.~Stump, J.~Huston, H.~L.~Lai, P.~Nadolsky and W.~K.~Tung,
  JHEP {\bf 0207}, 012 (2002)
  [arXiv:hep-ph/0201195].

\bibitem{Sjostrand:2003wg}
  T.~Sjostrand, L.~Lonnblad, S.~Mrenna and P.~Skands,
  arXiv:hep-ph/0308153.

\bibitem{pgs}
\url{http://www.physics.ucdavis.edu/~conway/research/software/pgs/pgs4-general.htm}

\bibitem{Gleisberg:2008fv}
  T.~Gleisberg and S.~H\"oche,
  arXiv:0808.3674 [hep-ph].

\bibitem{Duhr:2006iq}
  C.~Duhr, S.~H\"oche and F.~Maltoni,
  JHEP {\bf 0608}, 062 (2006)
  [arXiv:hep-ph/0607057].

\bibitem{Mangano:2002ea}
  M.~L.~Mangano, M.~Moretti, F.~Piccinini, R.~Pittau and A.~D.~Polosa,
  JHEP {\bf 0307}, 001 (2003)
  [arXiv:hep-ph/0206293].

\bibitem{Chivukula:1991zk}
  R.~S.~Chivukula, M.~Golden and E.~H.~Simmons,
  Nucl.\ Phys.\  B {\bf 363}, 83 (1991).

\bibitem{Dobrescu:2007yp}
  B.~A.~Dobrescu, K.~Kong and R.~Mahbubani,
  arXiv:0709.2378 [hep-ph].

\bibitem{Zerwekh:2008mn}
  A.~R.~Zerwekh, C.~O.~Dib and R.~Rosenfeld,
  Phys.\ Rev.\  D {\bf 77}, 097703 (2008)
  [arXiv:0802.4303 [hep-ph]].

\bibitem{Plehn:2008ae}
  T.~Plehn and T.~M.~P.~Tait,
  arXiv:0810.3919 [hep-ph].

\end{thebibliography}
\end{document}